\documentclass[a4paper,twocolumn,showpacs
,prb
,superscriptaddress
]{revtex4}

\usepackage{graphicx}
\usepackage{amssymb}
\usepackage{amsmath,amsfonts,latexsym}
\usepackage{array,tabularx}
\usepackage{dcolumn}               
\usepackage{color}

\DeclareMathOperator{\tr}{tr}
\DeclareMathOperator{\Tr}{Tr}

\DeclareMathOperator{\Ryx}{\mathcal{R}y_{x}}
\DeclareMathOperator{\ex}{x}
\DeclareMathOperator{\sech}{sech}

\newcommand{\vect}[1]{{\mathbf #1}}
\newcommand{\Frac}[2]{\displaystyle\frac{#1}{#2}}

\begin{document}

\title{Absorption, Photoluminescence and Resonant Rayleigh Scattering
Probes of Condensed Microcavity Polaritons}

\author{F.~M.~Marchetti} 
\affiliation{Cavendish Laboratory, University of Cambridge, Madingley
Road, Cambridge CB3 0HE, UK} 
\altaddress{Present address: Rudolf Peierls Centre for Theoretical
Physics, 1 Keble Road, Oxford OX1 3NP, UK}

\author{J.~Keeling}
\affiliation{Department of Physics, Massachusetts Institute of
  Technology, 77 Mass. Ave., Cambridge, MA 02139, USA}
\altaddress{Present address: Cavendish Laboratory, University of
Cambridge, Madingley Road, Cambridge CB3 0HE, UK}

\author{M.~H.~Szyma{\'n}ska}
\affiliation{Clarendon Laboratory, Department of Physics, University of Oxford,
             Parks Road, Oxford, OX1 3PU, UK}
\author{P.~B.~Littlewood}
\affiliation{Cavendish Laboratory, University of Cambridge,
             Madingley Road, Cambridge CB3 0HE, UK}

\date{Jan 13, 2006}       

\begin{abstract}
  We investigate and compare different optical probes of a condensed
  state of microcavity polaritons in expected experimental conditions
  of non-resonant pumping. We show that the energy- and
  momentum-resolved resonant Rayleigh signal provide a distinctive
  probe of condensation as compared to, e.g., photoluminescence
  emission. In particular, the presence of a collective sound mode
  both above and below the chemical potential can be observed, as well
  as features directly related to the density of states of
  particle-hole like excitations. Both resonant Rayleigh response and
  the absorption and photoluminescence, are affected by the presence
  of quantum well disorder, which introduces a distribution of
  oscillator strengths between quantum well excitons at a given energy
  and cavity photons at a given momentum. As we show, this
  distribution makes it important that in the condensed regime,
  scattering by disorder is taken into account to all orders.  We show
  that, in the low density linear limit, this approach correctly
  describes inhomogeneous broadening of polaritons. In addition, in
  this limit, we extract a linear blue-shift of the lower polariton
  versus density, with a coefficient determined by temperature and by
  a characteristic disorder length.
\end{abstract}


\maketitle

\section{Introduction}
\label{sec:intro}
Since the prediction of Keldysh and Kopaev,~\cite{keldysh} there has
been a long and intense pursuit to realise a condensed phase in solid
state excitonic, and related, systems. In particular, polaritons in
semiconductor microcavities, the coupled eigenstates of an exciton
with a cavity photon,~\cite{hopfield,weisbuch} represent ideal
candidates for observing condensation phenomena. The very light mass
of these composite bosonic particles promises relatively high
transition temperatures. In the past decade, improvements in the
growth technology of semiconductor heterostructures have made the
study of high-quality strongly-coupled planar microcavities almost
routine for III-V and II-VI semiconductors.  The high degree of
external control of these systems, and the possibility of direct
detection for them has opened the route towards a new generation of
fast optical matter-wave lasers and
amplifiers.~\cite{lesidang,bloch,exp_gen1,exp_gen2,exp_gen3} More
recently, concerted experimental efforts have been devoted to the
realisation of a Bose-Einstein condensate of microcavity
polaritons.~\cite{lesidang,deng,deng2,richard,richard_kavokin,hui_thermal,kasprzak06}

On the experimental side, a challenge to the realisation of a
condensed polariton phase might be represented by the finite quality
of the cavity mirrors and the resultant short polariton lifetime, of
the order of picoseconds. In addition, due to the `bottleneck
effect',~\cite{tassone} the relaxation of polaritons to the zero
momentum state can be delayed, hindering the creation of a thermal
population in the lowest energy state. It has however been recently
shown~\cite{tartakovskii,hui_thermal,kasprzak06} that thermalization
processes due to particle-particle scattering can be dramatically
magnified by increasing the value of the (non-resonant) pump power,
and by positively detuning the cavity energy above the excitonic
energy.  Under these conditions, the progress towards a zero momentum
quasi-equilibrium condensate has been
significant,~\cite{lesidang,bloch,deng,deng2,richard,richard_kavokin,hui_thermal}
including a non-linear threshold behaviour in the emission intensity
at zero momentum,~\cite{lesidang,bloch,deng,hui_thermal} the
investigation of the second order coherence function,~\cite{deng} a
characteristic change in the momentum space distribution above
threshold,~\cite{deng2,hui_thermal} and evidence that the
equilibration time is shorter than the polariton lifetime has been
seen.~\cite{hui_thermal} Finally, very recently, a clear demonstration
of condensation of cavity polaritons has been demonstrated in
CdTe.~\cite{kasprzak06} Kasprzak and collaborators have shown that
condensation of equilibrated polaritons can be achieved for effective
temperatures around $20$K and evidence for condensation has been seen
in the occupation function, the first order coherence (both in time
and in space) and in the spontaneous appearance of linear polarisation
of the condensate emission.

Alongside the experimental effort, a significant theoretical effort
has been invested in analysing properties and predicting signatures of
polariton
condensation.~\cite{paul,marzena,kavokin,kav_coh,jonathan,francesca,haug,kav_polaris,RRS_letter,marz_keldysh,porras,sarchi,quattropani}
Much of this work focuses on modelling the conditions under which
condensation can occur, both in
equilibrium,~\cite{paul,francesca,kav_coh,jonathan} and considering
the effects of pumping and
decay.~\cite{haug,marz_keldysh,sarchi,quattropani} Possible signatures
include the nonlinear relation of emission at zero momentum to pumping
power,~\cite{porras,sarchi,quattropani} changes to the
linewidth,~\cite{porras,quattropani,marz_keldysh} the PL spectrum and
the angular distribution of radiation,~\cite{jonathan,francesca} and
spontaneous polarisation of emitted radiation.~\cite{kav_polaris}

\begin{figure}
  \begin{center}
  \includegraphics[width=1\linewidth,angle=0]{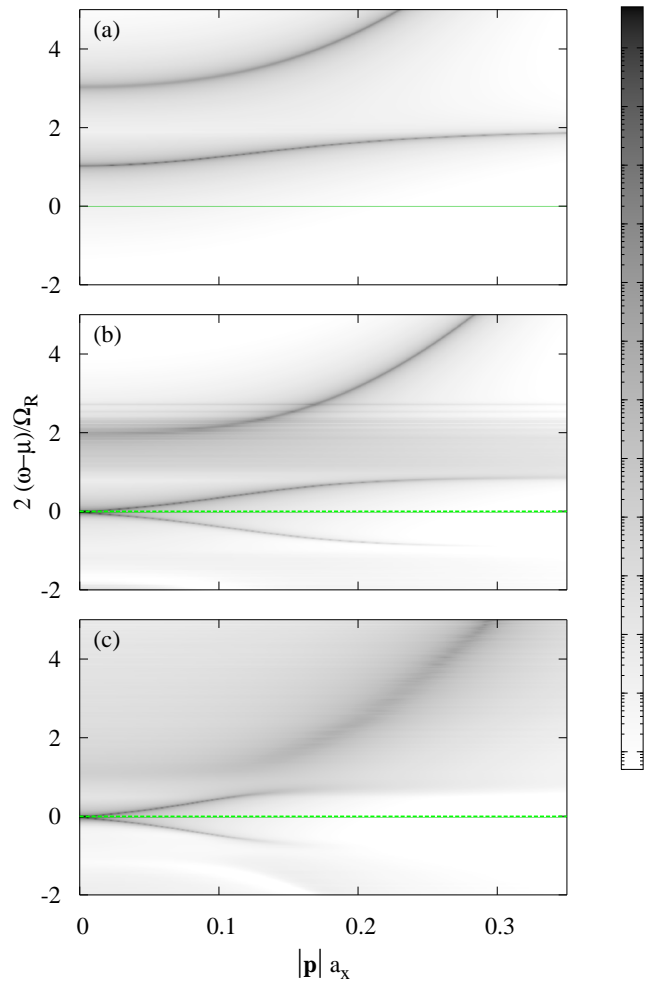}
  \end{center} \caption{\small (Color online) Contourplot of the
  disorder averaged RRS intensity $\langle I_{\vect{p} \vect{q}}
  (\omega)\rangle$ for $|\vect{p}| = |\vect{q}|$ as a function of the
  dimensionless momentum $|\vect{p}| a_{\ex}$ and rescaled energy
  $2(\omega - \mu)/\Omega_{\mathrm{R}}$, for zero detuning, Rabi
  splitting $\Omega_{\mathrm{R}}=26$meV, temperature $k_{B} T = 20$K,
  and a disorder strength characterize by an inverse scattering time
  $1/\tau = 1.16$meV. (a) non-condensed regime (dimensionless density
  $\rho\simeq 0$); (b) condensed regime ($\rho\simeq 7.8\times
  10^{-3}$); (c) condensed regime ($\rho\simeq 6.7\times 10^{-2}$).
  (The parameters chosen for these plots are the same as those used
  later for spectral weight and photoluminescence.)  The value of the
  chemical potential is explicitly marked [horizontal green (gray)
  line]. While in the non-condensed regime, RRS emission is always
  above the chemical potential, in the condensed phase, emission from
  the collective sound mode is seen both above and below the chemical
  potential.}
  \label{fig:rrsca}
\end{figure}

In this paper, we discuss the optical properties of condensed
polaritons, focusing our interests on absorption, photoluminescence
(PL) and resonant Rayleigh scattering (RRS). From our study we
conclude that RRS, the coherent scattering by disorder of polaritons
into directions other than that of the original probe, represents a
powerful tool for investigating the condensed phase.  We will show
that signatures of condensation are visible in the RRS spectrum,
allowing a direct probe of the collective excitation properties of the
polariton condensate.  In particular, we will show that, above the
threshold for condensation, strong emission from the collective sound
mode both above and below the chemical potential can be seen in the
RRS spectra (see Fig.~\ref{fig:rrsca}). In contrast such features are
expected to be much harder to observe in usual PL emission spectra,
where the spectrum is dominated by the very strong condensate emission
at the chemical potential, which is likely to mask these more subtle
features. In addition, we will show that the RRS spectra directly
reflects the disorder averaged density of states of excitonic
particle-hole like excitations, i.e. bound excitons coupled to the
coherent photon field.

Resonant Rayleigh scattering depends on disorder to scatter polaritons
between momentum states. In order to carry on our analysis, we
introduce a realistic description of disorder and analyse its effects
on RRS and more generally on other optical probes, like absorption and
PL. To do this, we make use of a quantitatively accurate model for
exciton
disorder,~\cite{zimmermann_poten,zimmermann_oscil,runge_review} and
numerically evaluate the distribution of excitonic energies and
oscillator strengths associated with a given disorder potential. An
accurate treatment of exciton disorder for quantum wells in
microcavities is important because the large ratio of exciton mass to
photon mass means that those exciton states contributing most to the
thermally populated polaritons are strongly localised, i.e. cannot be
treated within the first Born approximation. Such a treatment shows
that, for a given exciton energy and momentum, there is still a
distribution of possible oscillator strengths.  For the exciton states
relevant to polariton formation, the distribution of oscillator
strengths varies from a narrow Gaussian at low energies (below the
band edge) to a Porter-Thomas distribution at high energies, as is
known from previous
works.~\cite{zimmermann_poten,zimmermann_oscil,runge_review} Here, we
study the effect of this distribution on the many-body physics, and in
particular, on RRS and PL spectra.

In this paper, we consider the case relevant to the existing CdTe or
GaAs microcavities of high-quality quantum wells, where the typical
excitonic disorder amplitude is smaller than the Rabi splitting. In
this regime, we can make use of the \emph{coupled oscillator model} or
wave vector conserving approximation, as explained in
Ref.~\onlinecite{whittaker}.  As will be explained in more detail
later on, this corresponds to approximating the full exciton Green's
function by its momentum-diagonal part (or equivalently considering
its disorder average), and treating perturbatively the off-diagonal
terms. In this way, the translational invariance broken by the
presence of the disorder is restored, polaritons have a defined wave
vector and, at high enough densities, condense in the lowest momentum
state. Such a treatment implies that excitonic disorder, being on
short length scales, does not lead to spatial inhomogeneity on the
length scale associated with the polariton. Extended polaritons are
formed from a superposition of many localised excitons, therefore
recovering translational invariance at the level of polaritons. At low
densities, in the non-condensed state, many observable properties can
be adequately described by the coupled oscillator model and so can be
found from the mean squared oscillator strength at a given energy.  In
this limit our method recovers the well known results for the
inhomogeneous broadening of the polariton PL,~\cite{whittaker,borri}
and --- by considering in addition the mean fourth power of the
oscillator strength --- the averaged RRS
response.~\cite{whittakerRRS,shchegrov,freixanet,langbein} However,
when condensed, there are observable effects associated with the full
distribution of oscillator strengths. In particular, the non-vanishing
probability of excitons to have arbitrarily small oscillator strengths
has direct consequences for optical probes, including both RRS and PL.

In Ref.~\onlinecite{RRS_letter}, we considered specific aspects of
resonant Rayleigh scattering arising from the model discussed in this
paper. Here, we provide and compare further experimental probes of
condensation, and discuss the underlying physical mechanisms involved.
All these optical probes, including RRS, are in addition to a
non-resonant pumping of the microcavity polaritons. A related problem
is studied in Ref.~\onlinecite{carusotto}. There, the Rayleigh
scattering of a strong resonant pump is considered (treating disorder
perturbatively), thus the probe is the pump. In this paper, in
contrast, one has to distinguish two types of coherence. The first
type is the internal coherence of the condensed system, which arises
spontaneously following non-resonant pumping. The second type is the
coherent scattering of an external laser probe on disorder, which can
be strongly modified by the presence of the condensate.

The form of the density of states, optical density, and distribution
of oscillator strengths found from the numerical calculations are
also important in other thermodynamic properties and probes of
polaritons. One such example is the calculation of the linear
blue-shift vs. density in the low density regime. The inclusion of
exciton states that couple weakly to photons extends the validity of
the current model to densities beyond those in similar
treatments,~\cite{paul,jonathan} and provides a stronger basis for the
use of those models in the regimes where they are valid.

The paper is arranged in the following sections: We describe the
effect of quantum well disorder on the exciton oscillator strength in
Sec.~\ref{sec:disor}. There, we also give details of the numerical
analysis (Sec.~\ref{sec:numer}), which will be used in the many-body
model introduced in Sec.~\ref{sec:model} to describe the
thermodynamics of polariton condensation
(Sec.~\ref{sec:phase}). Section~\ref{sec:photo} describes the optical
probes such as spectral weight and photoluminescence
(Sec.~\ref{sec:spect}) and resonant Rayleigh scattering
(Sec.~\ref{sec:rrsca}). Conclusions are collected in
Sec.~\ref{sec:concl}, while Appendix~\ref{sec:phase-sens-detect}
explains how one can detect RRS using phase sensitive measurements.

\section{Excitons in Disordered Quantum Wells}
\label{sec:disor}
The problem of an exciton in a disordered quantum well has been
studied at some length in the last two
decades.~\cite{efros,efros2,agranovich,zimmermann_poten,zimmermann_oscil,runge_review,savona_review}
Quantum well disorder can arise due to interface and alloy
fluctuations and affects the properties of the excitonic linewidth and
absorption spectrum. Similarly to
Refs.~\onlinecite{zimmermann_poten,zimmermann_oscil}, we will assume
the external disorder potential to be correlated on a length scale
larger than the exciton Bohr radius. Accordingly, we factorise the
excitonic in-plane relative and centre of mass coordinates
\begin{align*}
  \Psi_{\alpha} (\vect{r}_e, \vect{r}_h) &\simeq \varphi_{1s} (r)
  \Phi_{\alpha} (\vect{R}) \\
  \varphi_{1s} (r) &= \sqrt{8/\pi a_{\ex}^2} e^{-2r/a_{\ex}}\; ,
\end{align*}
and assume that the disorder affects only the excitonic centre of mass
motion $\Phi_{\alpha} (\vect{R})$, while the internal degrees of
freedom can be restricted to the ground state hydrogenic state
$\varphi_{1s} (r)$. Here, $a_{\ex} = \epsilon/e^2 \mu_r$ is the
exciton Bohr radius and $\mu_r$ the reduced mass (henceforth we will
set $\hbar = 1$). Neglecting the transverse degrees of freedom related
to the confinement of the excitons in the QW, the energy associated to
the wavefunction $\Psi_{\alpha} (\vect{r}_e, \vect{r}_h)$ is given by
the sum of the relative motion energy $E_{\ex}$ [i.e., the band gap
minus the exciton binding energy, $\Ryx = (2\mu_r a_{\ex}^2)^{-1} =
e^4 \mu_r/2\epsilon^2$] and the energy related to the centre of mass
motion $\varepsilon_{\alpha}$:
\begin{equation}
  \left[-\Frac{\nabla_{\vect{R}}^2}{2m_{\ex}} + V(\vect{R}) +
  E_{\ex}\right] \Phi_{\alpha} (\vect{R}) = \varepsilon_{\alpha}
  \Phi_{\alpha} (\vect{R})\; .
\label{eq:centr}
\end{equation}
Here, the effective disorder potential $V (\vect{R})$ represents the
microscopic structural disorder averaged over the electron-hole
motion.~\cite{zimmermann_poten} This can be approximated, e.g., with a
Gaussian noise correlated on a length scale $\ell_c>a_{\ex}$ with
variance equal to $\sigma^2$:
\begin{equation*}
  \langle V(\vect{R}) V(\vect{R}')\rangle = (\sigma^2 \ell_c^2/L^2)
  \sum_{\vect{q}}^{1/\ell_c} e^{i\vect{q} \cdot (\vect{R} -
  \vect{R}')}\; ,
\end{equation*}
where $L^2$ is the quantisation area. It will be convenient, later on,
to introduce the scattering time
\begin{equation*}
   \tau=\Frac{1}{2\pi \nu \sigma^2 \ell_c^2} \; ,
\end{equation*}
where $\nu = m_{\ex}/2\pi$ is the two-dimensional density of states in
the clean limit.

In two dimensional non-interacting systems, all states are localised
by the disorder potential. However, the localisation length and the
character of the excitonic wavefunction change significantly from
below to above the band edge $E_{\ex}$. Well below the band edge,
low-energy Lifshitz tail states,~\cite{lifshitz} are well localised in
deep potential minima, with a nodeless (roughly Gaussian) shape. These
states are rare, because they occur due to large, rare, fluctuations
of the disorder potential.  In contrast, higher energy states above
the band edge have a fractal-like shape with many nodes and can be
approximated by a random superposition of plane waves with the same
momentum $|\vect{p}| \simeq \sqrt{2m_{\ex}
(\varepsilon_{\alpha}-E_{\ex})}$. Here, the localisation mechanism is
closely related to quantum mechanical interference effects.

Accordingly, the change of the shape of the centre of mass
wavefunctions across the band edge is accompanied by a dramatic change
in the excitonic oscillator strength.~\cite{zimmermann_oscil} The
oscillator strength describing coupling of a quantum well exciton to
light, $g_{\alpha,\vect{p}}$, is given by the probability amplitude of
finding an electron and a hole at the same position and with centre of
mass momentum equal to the photon momentum $\vect{p}$ and is therefore
proportional to the Fourier transform of the centre of mass wave
function, $\Phi_{\alpha,\vect{p}} = \langle \Phi_{\alpha} |
\vect{p}\rangle$:
\begin{equation}
  g_{\alpha ,\vect{p}} = e d_{ab} \sqrt{\Frac{2\pi
  \omega_{\vect{p}}}{\epsilon L_w}} \varphi_{1s} (0)
  \Phi_{\alpha,\vect{p}}\; ,
\label{eq:oscil}
\end{equation}
where, $d_{ab}$ is the dipole matrix element. The dispersion for
photons in a microcavity of width $L_w$ is given by $\omega_{\vect{p}}
= \sqrt{\omega_0^2 + (c\vect{p})^2/\epsilon}$, where $\omega_0 = 2 \pi
c /L_w \sqrt{\epsilon}$, and can be approximated, for small momenta,
by a parabolic dispersion $\omega_{\vect{p}} \simeq \omega_0 +
\vect{p^2}/2 m_{ph}$, with the photon mass given by $m_{\text{ph}} =
2\pi\sqrt{\epsilon}/cL_w$.

Because there may be many different exciton wavefunctions
corresponding to similar exciton energies, the oscillator strength
$g_{\alpha ,\vect{p}}$ is a random quantity, which varies both in
phase and magnitude. Considering many disorder realisations, we find a
distribution of squared oscillator strengths, $|g_{\alpha
,\vect{p}}|^2$ as a function of the energy $\varepsilon_{\alpha}$ and
momentum $\vect{p}$.  This distribution reflects the statistical
properties of the centre of mass excitonic wave functions.

To compare to experimentally relevant observables, we introduce the
density of states (DoS),
\begin{equation*}
  \text{DoS} (\varepsilon) \equiv \Frac{1}{L^2} \sum_{\alpha} \langle
\delta (\varepsilon - \varepsilon_{\alpha})\rangle \; ,
\end{equation*}
and the mean squared oscillator strength:
\begin{equation}
  g^2(\varepsilon,|\vect{p}|) = \Frac{1}{\text{DoS}
  (\varepsilon)}\langle \sum_{\alpha} |g_{\alpha,\vect{p}}|^2
  \delta(\varepsilon - \varepsilon_\alpha)\rangle \; ,
\label{eq:avgep}
\end{equation}
where $\langle \dots \rangle$ is the average over different disorder
realisations. These quantities are related to the excitonic optical
density by the relation:
\begin{equation}
  D(\varepsilon) = \text{DoS} (\varepsilon) g^2 (\varepsilon,0)\; .
\label{eq:optde}
\end{equation}
Experimentally, the excitonic optical density can be measured by
dividing the PL emission by the excitonic occupation.~\cite{borri}

\begin{figure}
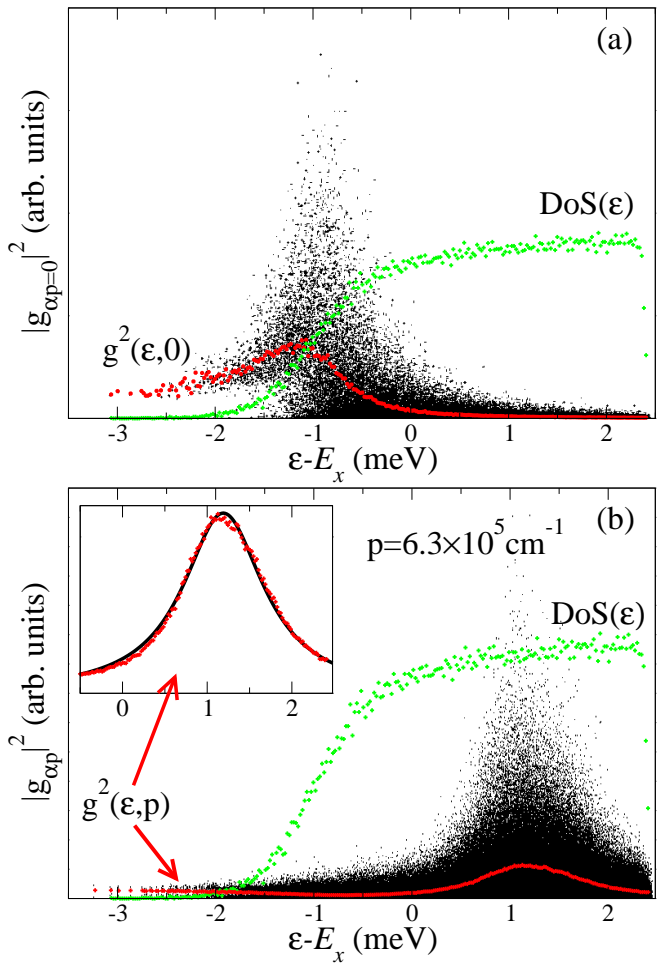

  \begin{center}
    \includegraphics[width=1\linewidth,angle=0]{allg_modp0_vs2.eps}

    \includegraphics[width=1\linewidth,angle=0]{allg_modp10_vs4.eps}
  \end{center}
  \caption{\small (Color online) Plot showing the energy dependence of
    the excitonic squared coupling strength $|g_{\alpha,\vect{p}}|^2$
    to photons of momentum $|\vect{p}| = 0$ (a) (from
    Ref.~\onlinecite{RRS_letter}, for comparison) and $|\vect{p}| =
    6.3 \times 10^{5} \mathrm{cm}^{-1}$ (b), where all exciton states
    found numerically. Results are taken from 160 different
    realizations of disorder potential, and for (b), coupling
    strengths from eight different photon momenta $(p_x,p_y)$ with the
    same value of $|\vect{p}|$ are combined. The mean squared averaged
    oscillator strength $g^2 (\varepsilon,|\vect{p}|)$ for the same
    value of momentum [lower red (gray) points] and the density of
    states $\text{DoS} (\varepsilon)$ [upper green (gray) points] are
    also explicitly plotted. Inset: Fit of $g^2
    (\varepsilon,|\vect{p}|)$ to the expression~\eqref{eq:borna} with
    a renormalized energy $\varepsilon_{\vect{p}}$.}
  \label{fig:gepsp}
\end{figure}

Focusing for the moment on average properties, rather than on the
entire distribution, a simple expression exists for the mean squared
oscillator strength in the high energy limit. In this limit, where the
DoS is flat and energy independent, $\text{DoS} (\varepsilon) \simeq
\nu$, making use of the Born approximation, the squared average
oscillator strength $g^2(\varepsilon,|\vect{p}|)$ does not depend
separately on the disorder potential correlation length $\ell_c$ and
variance $\sigma$, but instead only on the scattering time $\tau$ (see
Fig.~\ref{fig:gepsp}):
\begin{equation}
  g^2(\varepsilon,|\vect{p}|) \simeq \Frac{1}{m_{\ex} \tau L^2}
  \Frac{1}{(\varepsilon - \varepsilon_{\vect{p}})^2 + (1/2\tau)^2} \;
  ,
\label{eq:borna}
\end{equation}
where $\varepsilon_{\vect{p}}=E_{\ex} + \vect{p}^2/2m_{\ex}$ is the
free particle dispersion.  From this form, and the effectively
constant density of states, one can see that both the mean squared
oscillator strength and the excitonic optical density are symmetrical.
The comparison of this approximate form to the numerical simulation is
shown in Fig.~\ref{fig:gepsp},

In contrast, for energies much below the band edge, the specific
asymptotic expression of $g^2(\varepsilon,|\vect{p}|)$ depends whether
the correlation length $\ell_c$ is smaller or larger than the
localisation length $r_{\Phi} (\varepsilon)$. In the white noise
limit,\cite{lifshitz} $\ell_c \ll r_{\Phi} (\varepsilon) \sim (2
m_{\ex} |\varepsilon - E_{\ex}|)^{-1/2}$, one can show that the centre
of mass wavefunction $\Phi_{\alpha} (\vect{R})$ can be approximated by
a Gaussian centred at a randomly distributed site $\vect{R}_{\alpha}$,
\begin{align*}
  \Phi_{\alpha} (\vect{R}) &= r_{\Phi}^{-1} e^{-(\vect{R} -
  \vect{R}_{\alpha})^2/r_{\Phi}^2}\\ \Phi_{\alpha,\vect{p}} &=
  r_{\Phi} e^{i\vect{p} \cdot \vect{R}_{\alpha} - (r_{\Phi}
  \vect{p})^2/4}\; ,
\end{align*}
and therefore giving a squared oscillator strength proportional to:
\begin{equation*}
  |\Phi_{\alpha,\vect{p}}|^2 \simeq \Frac{1}{2 m_{\ex} |\varepsilon -
  E_{\ex}|} e^{-|\vect{p}|^2/(4m_{\ex} |\varepsilon - E_{\ex}|)} \; .
\end{equation*}
Thus, here the distribution of squared oscillator strengths is very
narrow, with a mean square value given by the above form.

Similarly, much theoretical (and numerical) work has been done to
establish the energy dependence of the density of states.  In the low
energy tail, as before, the specific asymptotic form of the DoS
depends on the value of the correlation length $\ell_c$. In the white
noise limit, one can show that $\text{DoS} (\varepsilon) \propto
|\varepsilon - E_{\ex}|^{3/2} e^{-11.8 |\varepsilon -
E_{\ex}|/\sigma^2 m_{\ex} \ell_c^2}$, while, in the opposite
(classical) limit, $\ell_c \gg r_{\Phi}' (\varepsilon) \sim (2 \ell_c
/\sqrt{2m_{\ex} |\varepsilon - E_{\ex}|})^{1/2}$, instead one has
$\text{DoS} (\varepsilon) \propto (\varepsilon - E_{\ex})^{2} e^{-
(\varepsilon - E_{\ex})^2/2 \sigma^2}$ (see, e.g.,
Ref.~\onlinecite{agranovich}). In general, for the finite values of
the disorder correlation length corresponding to typical experiments,
the regions in energy where one of these two analytical regimes apply
are very restricted, and therefore a numerical analysis is required.

Numerical analysis is also essential in order to account for the
distribution of squared oscillator strengths $|g_{\alpha,\vect{p}}|^2$
near the band edge. This distribution changes substantially from low
energy states to high energy ones. As we discuss more in detail in the
next section, for high energy states the oscillator strength
distribution is governed by a Porter-Thomas law, while for Lifshitz
tail states the distribution follows a narrow Gaussian-like
distribution. Neither of these distributions apply for energies around
the band edge. However, it is states near the band edge that have the
largest optical density, and so the distribution of oscillator
strengths in this region has a significant impact on derived
quantities. It thus becomes essential to use numerical analysis to
find the entire distribution of oscillator strengths.

\begin{figure}
  \begin{center}
    \includegraphics[width=1\linewidth,angle=0]{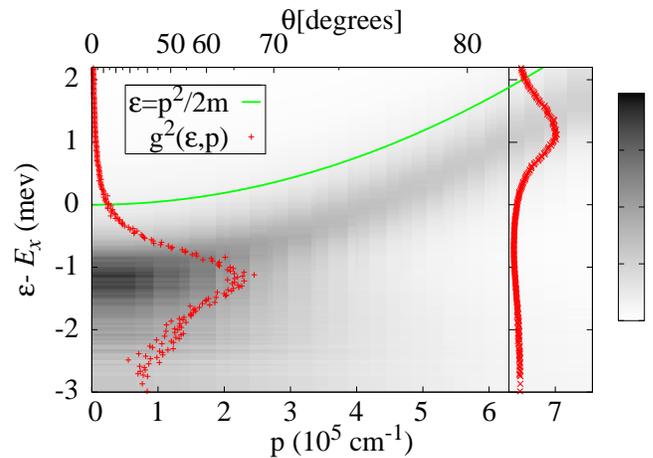}
  \end{center}
  \caption{\small (Color online) Contourplot of the mean (over 160
    realizations of the disorder potential) squared oscillator
    strength $g^2 (\varepsilon,|\vect{p}|)$ versus energy and momentum
    [or equivalently angle $\theta = \tan^{-1} (c
    |\vect{p}|/\omega_0)$].  Note that the scale in angle is not
    linear. The free particle dispersion $E_{\ex} +
    |\vect{p}|^2/2m_{\ex}$ [solid green (gray) line], and the a trace
    of the mean squared oscillator strength for two representative
    values of momenta, $|\vect{p}|=0$ ($\theta = 0^{\circ}$) and
    $|\vect{p}|=6.3 \times 10^{5} \mathrm{cm}^{-1}$ ($\theta =
    82^{\circ}$) [red (gray) plus symbols] are explicitly plotted (cf
    Fig.~\ref{fig:gepsp}). The figure is adapted from
    Ref.~\onlinecite{RRS_letter}.}
  \label{fig:geps2}
\end{figure}
%

\subsection{Numerical Analysis}
\label{sec:numer}
\begin{figure}
\begin{center}
\includegraphics[width=1\linewidth,angle=0]{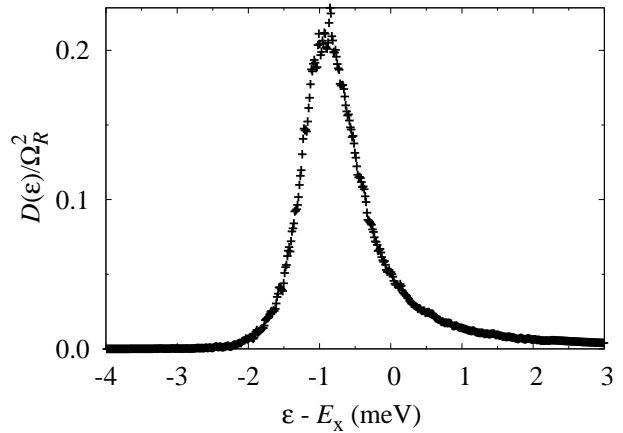}
\end{center}
\caption{\small Optical density Eq.~\eqref{eq:optde} versus
         energy. The maximum value is around $\varepsilon^* -
         E_{\ex}\simeq -0.94$meV and the FWHM $\sigma^*\simeq0.94$meV.}
\label{fig:optic}
\end{figure}

To solve Eq.~\eqref{eq:centr}, we exactly diagonalise this problem
within a finite basis set, using conjugate-gradient minimisation with
pre-conditioning of the steepest descent wavevector (for a detailed
discussion of this method, see Ref.~\onlinecite{payne}).  We find the
exact eigenvalues and eigenfunctions in a finite basis on a grid of
$120 \times 120$ points for a system of size $L=1\mu$m, $\sigma=2$meV,
$\ell_c=166$\AA\ and excitonic mass $m_{\ex}/m_{0}=0.08$. These
parameters give an inverse scattering time of $1/\tau=1.16$meV. For
this choice of the grid, one can show that convergence is reached.
From the evaluated eigenvalues $\varepsilon_\alpha$ and eigenstates
$\Phi_{\alpha,\vect{p}}$ over 160 realisations of the disorder
potential, we can derive the excitonic density of states, the
oscillator coupling strength and its squared average,
Eq.~\eqref{eq:avgep}, which we plot in Figs.~\ref{fig:gepsp},
and~\ref{fig:geps2}, while the corresponding optical density is
plotted in Fig.~\ref{fig:optic}.

The lower panel of Fig.~\ref{fig:gepsp} shows the squared coupling
strength $|g_{\alpha,\vect{p}}|^2$ versus energy for a fixed value of
momentum, $|\vect{p}| = 6.3 \times 10^{5} \mathrm{cm}^{-1}$,
corresponding, for a cavity of $\omega_0=1.68$eV, to an angle of
$\theta = \tan^{-1} (c |\vect{p}|/\omega_0)= 82^{\circ}$. Note that,
because of the presence of disorder in the quantum well, one photon
with a given momentum couples with many exciton states with different
energies. We will see later on therefore that a polariton with a given
momentum is formed by the superposition of one photon state
$|\vect{p}\rangle$ and many exciton states $| \Phi_{\alpha} \rangle$.
These states are more or less strongly coupled depending on the
distribution of oscillator strength for that given momentum.
As Fig.~\ref{fig:gepsp} shows, by probing a quantum well at a large
angle, i.e. with high momenta photons, the excited excitons with
larger oscillator strength are the ones which are almost-delocalised
in nature and with a many-node fractal-like shape.  By taking the
average over many (160) disorder realisations, the squared average
oscillator strength $g^2 (\varepsilon,|\vect{p}|)$ shown in the inset
in Fig.~\ref{fig:gepsp} is well described by the Lorentzian shape
predicted by the Born approximation of Eq.~\eqref{eq:borna}, with a
fitted width of $1/\tau_{\text{fit}} \simeq 1.2$meV in good agreement
with the theoretical value $1/\tau=1.16$meV. However, the peak of the
Lorentzian does not coincide with the energy of the clean limit,
$\varepsilon_{\vect{p}}$, but is renormalised down in value, as can be
shown by employing a self-consistent Born approximation.

In contrast, as shown in the upper panel of Fig.~\ref{fig:gepsp}, for
photons with zero momentum, the maximum value of the oscillator
strength characterises excitonic states below the band edge which are
more localised in nature. This can be easily understood by the
following qualitative argument: At very low energies, the excitonic
state is strongly localised in a deep potential minimum and has no
nodes.  Increasing the energy, at first the localisation length
increases [e.g., $r_{\Phi} (\varepsilon) \sim (2m_{\ex} |\varepsilon -
E_{\ex}|)^{-1/2}$ in the white noise limit] and thus increases the
oscillator strength. However, eventually, the wavefunction starts
developing nodes and consequently the squared average oscillator
strength decreases. When $\vect{p}=0$, only the high energy side of
$g^2 (\varepsilon,0)$ can be described by the Born
limit~\eqref{eq:borna}.  For the chosen values of the disorder
potential correlation length $\ell_c$ and variance $\sigma$, an
analytical expression for the low energy (Lifshitz) tail is not known.

The crossover from localised to more plane-wave-like excitonic states
obtained by increasing the value of the photon momentum is plotted in
Fig.~\ref{fig:geps2}.  At high momenta, the maximum of the squared
average oscillator strength $g^2 (\varepsilon,|\vect{p}|)$ shown in
Fig.~\ref{fig:geps2} follows the free particle dispersion, plus a
renormalisation down in energy, which decreases for higher photon
momenta, i.e. these states are described well by free particles,
including disorder in the first Born approximation. At low momenta,
the states which have the stronger oscillator strength are effectively
localised. The crossover, as seen in the contour plot, happens at
relatively large angle, $\theta \simeq 55^{\circ}$, because of the
large ratio of exciton to photon mass: The crossover momentum is set
by exciton mass, but its conversion to an angle depends on effective
photon mass. For a connected reason, the thermally populated polariton
states are formed out of strongly localised (i.e. beyond first Born
approximation) excitonic states: For values of temperature and photon
mass relevant for experiments, thermal population of polaritons
extends up to around $10^{\circ}$ in momentum, which as seen in
Fig.~\ref{fig:geps2} corresponds to exciton states not accurately
described by the Born approximation. For this reason, in the following
we will concentrate on the oscillator strength corresponding to
$\vect{p} \simeq 0$, $|g_{\alpha,0}|^2$.

Finally we plot in Fig.~\ref{fig:optic} the optical density. For the
chosen values of $\ell_c$ and $\sigma$, the optical density shows a
maximum around $\varepsilon^* - E_{\ex}\simeq -0.94$meV below the band
edge, a full width at half maximum (FWHM) of approximatively
$\sigma^*\simeq0.94$meV and a clear asymmetry of the line-shape. It is
well known~\cite{runge_review,savona_review} that, by indicating with
$E_c=1/2m_{\ex }\ell_c^2$ the confinement energy of the lowest state
in a typical potential minimum, the excitonic line shape is determined
by the ratio $E_c/\sigma$ and, for a finite value of this, the optical
density develops an asymmetry towards higher energies. For our choices
of parameters, $E_c/\sigma \simeq 0.85$. Asymmetry of the optical
density of quantum well excitons has been also measured experimentally
(see, e.g., Ref.~\onlinecite{borri}), by dividing the measured PL by
the excitonic occupation.

As mentioned earlier, the full distribution of oscillator strengths,
and not just its mean squared value will be important.  It is useful
to discuss here some technical details of how this is extracted. The
numerical analysis provides the excitonic eigenvalues and eigenstates
only within a finite interval in energy.  (The lowest energy states
arise from rare potential fluctuations, which would require a larger
region of space to be sampled; the highest energy states have spatial
variation on lengthscales finer than our grid.)  Within the interval
of energies found, averages are performed making use of the raw data
coming from 160 realisations of the disorder potential;  outside this
interval, averages are taken by extrapolating the numerics. In
particular, in the low energy Lifshitz tail, as the oscillator
strengths have a narrow Gaussian distribution around its squared
averaged value, we approximate the distribution of $|g_{\alpha,0}|^2$
with a delta function at its extrapolated value $g^2(\varepsilon,0)$.
In the very high energy region, instead, we make use of the
Porter-Thomas distribution,
\begin{equation*}
  \mathcal{P} (x=|g_{\alpha,0}|^2) = \Frac{\exp[-x/(2
  \bar{x})]}{\sqrt{2 \pi x \bar{x}}} \; ,
\end{equation*}
where $\bar{x}=g^2(\varepsilon,0)$, again extrapolating the fitted
value for the squared averaged oscillator strength. In addition, we
fix the overall scale of $|g_{\alpha,0}|^2$ to match integrated
optical intensity with half the experimentally measured Rabi splitting
$\Omega_R$ squared:
\begin{equation}
  \int d\varepsilon D(\varepsilon) = \Frac{\Omega_R^2}{4}
\label{eq:fixgi}
\end{equation}
This normalisation accounts for factors other than the wavefunction in
Eq.~(\ref{eq:oscil}).

Having described the effect of quantum well disorder on the energies
of excitons and on their dipole coupling to the cavity photons, we
will next introduce the many-body Hamiltonian that will allow us to
describe a system of thermalized interacting polaritons which, for a
given density and temperature, can condense to a superfluid phase.

\section{The Polariton Model}
\label{sec:model}
The Hamiltonian describing a model of excitons with random energies
$\varepsilon_{\alpha}$, corresponding to the centre of mass eigenstate
from Eq.~\eqref{eq:centr}, dipole coupled to a cavity mode can be
written as
\begin{multline}
  \hat{H} = \sum_{\alpha} \Frac{\varepsilon_{\alpha}}{2}
  \left(b_{\alpha}^\dag b_{\alpha} + a_{\alpha} a_{\alpha}^\dag\right)
  + \sum_{\vect{p}} \omega_{\vect{p}} \psi_{\vect{p}}^\dag
  \psi_{\vect{p}} \\ + \Frac{1}{\sqrt{L^2}} \sum_{\alpha}
  \sum_{\vect{p}} \left(g_{\alpha , \vect{p}} \psi_{\vect{p}}
  b_{\alpha}^\dag a_{\alpha} + \text{h.c.}\right) \; .
\label{eq:hamil}
\end{multline}
The description of exciton-exciton interaction in this model is
somewhat simplified: it imposes the constraint that two excitons
cannot occupy the same energy level $\varepsilon_{\alpha}$, because of
their composite fermionic nature.  This constraint is implemented by
representing each energy level [where energy levels are distributed
according to the density of states, $\text{DoS} (\varepsilon)$] by two
possible states, respectively $|\text{g.s.} \rangle$ and $|\text{ex.}
\rangle$.  These two states are then represented by occupation of two
fermionic states, so that the ground state $|\text{g.s.}\rangle =
a_{\alpha}^\dag |0\rangle$, and an excitonic state $|\text{ex.}
\rangle = b_{\alpha}^\dag |0\rangle = b_{\alpha}^\dag a_{\alpha}
|\text{g.s.}\rangle$.  Imposing a constraint on total fermion
occupancy, $b_{\alpha}^\dag b_{\alpha} + a_{\alpha}^\dag a_{\alpha} =
1$, eliminates the unphysical states $|0\rangle$ and $a_{\alpha}^\dag
b_{\alpha}^\dag |0\rangle$.  In this way saturation effects are
introduced to all orders.  Note that these fermionic operators
$a^{\dagger}_{\alpha}$ cannot be written as a linear combination of
free electron and hole operators -- this is a result of having already
included the intra-exciton Coulomb interaction.

The transition matrix element between a photon state
$\psi_{\vect{p}}^\dag |\text{g.s.} \rangle$ and an exciton state,
$\langle \text{ex.}| = \langle \text{g.s.}| a_{\alpha}^\dag
b_{\alpha}$, is given by $g_{\alpha,\vect{p}}$ from
Eq.~\eqref{eq:oscil} and evaluated numerically, as explained in the
previous section.  It is convenient to rescale this coupling according
to
\begin{equation*}
  g_{\alpha , \vect{p}} \mapsto g_{\alpha , \vect{p}} \sqrt{\Ryx
  m_{\ex}/2\pi}\; ,
\end{equation*}
where $N = \Ryx L^2 m_{\ex}/2\pi$ is the inverse level spacing $L^2
m_{\ex}/2\pi$, measured in units of the excitonic Rydberg energy. This
corresponds to measuring the density of particles per Bohr radius
squared. Using these units, we may write the total number of
excitations
\begin{equation}
  \hat{N} = \sum_{\alpha} \Frac{1}{2} \left(b_{\alpha}^\dag b_{\alpha}
  + a_{\alpha} a_{\alpha}^\dag\right) + \sum_{\vect{p}}
  \psi_{\vect{p}}^\dag \psi_{\vect{p}}
\label{eq:numbe}
\end{equation}
n a dimensionless form by introducing a dimensionless density of
particles $\rho \equiv \langle \hat{N} \rangle /N$, or equivalently
$\rho= (\langle \hat{N}\rangle/L^2) a_{\ex}^2\, 4\pi \mu/m_{\ex}$,
where $\langle \hat{N}\rangle/L^2$ is the physical areal density of
particles.

\begin{figure}
\begin{center}
\includegraphics[width=0.9\linewidth,angle=0]{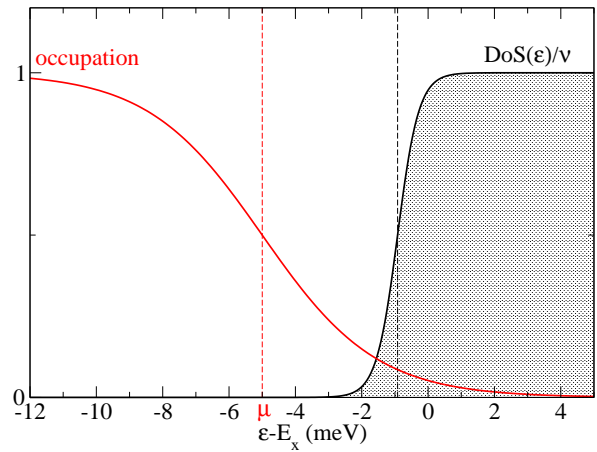}
\end{center}
\caption{\small (Color online) Normalized density of states
  $\text{DoS} (\varepsilon)/\nu$ and thermal occupation factor [red
  (gray) line starting at top left] for $E_{\ex} - \mu = 5$meV and
  $k_BT=20$K.}
\label{fig:occup}
\end{figure}

In the limit of vanishing density, our occupancy constraint has no
effect, and so this description is equivalent to one treating excitons
as bosons. As will be shown later, in Sec.~\ref{sec:blues}, for low
densities, our occupancy constraint is also equivalent to a model of
excitons as bosons, but with some effective saturation
interaction.~\footnote{Formally this can be seen by representing the
two levels as a spin 1/2, and making a Holstein-Primakoff
transformation of our model.} Thus, as in the bosonic case, saturation
effects are linear~\cite{rochat,ciuti} and the blue-shift increases
linearly in density. There is however an important difference to those
previous works. In contrast to the clean systems studied there, our
effective saturation interaction has a coefficient related to disorder
and temperature rather than to the Rabi splitting and Bohr radius
only. However, increasing the density, saturation effects, and thus
the underlying fermionic degrees of freedom become more and more
important, and the conventional linear treatment becomes
inadequate. In our treatment saturation effects are introduced exactly
and taken into account to all orders. Our model is accurate as far as
the occupancy is restricted to the strongly localised Lifshitz states
in the tail below the band edge (see Fig.~\ref{fig:occup}). All the
results described here respect this limit. One can moreover show that
it is this density regime which is relevant for on-going experiments
in CdTe.~\cite{kasprzak06} Higher energy states, beyond the Lifshitz
tail, contribute to the optical response, however they are only weakly
occupied, and so the above limit remains valid.

What the model Eq.~\eqref{eq:hamil} does not include is the Coulomb
interaction between excitons with different energies
$\varepsilon_{\alpha}$, i.e., at low densities, on different
localisation sites $\vect{R}_{\alpha}$. This contribution is expected
to be small in the low density regime. This low density regime will be
observed all the way through this paper. At the same time we are not
including double-occupancy of a single exciton energy level, which
could be important at higher densities.

In the following, we will make extensive use of the numerical analysis
of oscillator strength distributions explained in the previous
section.  Let us note here again that in the non-condensed regime, the
full distribution is unnecessary -- the excitonic optical density is
sufficient to derive the polariton dispersion and its inhomogeneous
broadening.  However, in the condensed phase, and in order to
correctly describe resonant Rayleigh scattering mechanisms, we will
see that it is of fundamental importance to consider the entire
distribution of oscillator strength.

We consider a thermal equilibrium system of polaritons, as describing
the situation with non-resonant pumping and strong thermalization
processes. This limit has been recently demonstrated to be accessible
in on-going experiments.~\cite{hui_thermal,kasprzak06} The total
density of polaritons is fixed by introducing a chemical potential
$\mu$.  Making use of standard path integral techniques and of the
grand canonical ensemble, $\hat{H} - \mu \hat{N}$, we integrate over
the fermionic fields, and thus write the partition function
$\mathcal{Z} = \tr [e^{-\beta (\hat{H} -\mu \hat{N})}]$, with
$\beta=1/k_B T$, in terms of the following imaginary time action:
\begin{equation}
  S[\psi] = \int_0^\beta d\tau \sum_{\vect{p}} \psi_{\vect{p}}^*
  \left(\partial_\tau + \tilde{\omega}_{\vect{p}}\right)
  \psi_{\vect{p}} - \Tr \ln G^{-1} \; .
\label{eq:actio}
\end{equation}
Here, $G^{-1}$ is the, energy level diagonal, inverse single-particle
Green's function:
\begin{equation}
  G^{-1}_{\alpha} =
  \begin{pmatrix} 
    \partial_\tau + \tilde{\varepsilon}_{\alpha}/2 & \sum_{\vect{p}}
    g_{\alpha , \vect{p}} \psi_{\vect{p}}/\sqrt{N}\\ \sum_{\vect{p}}
    g_{\alpha , \vect{p}}^* \psi_{\vect{p}}^*/\sqrt{N} & \partial_\tau
    - \tilde{\varepsilon}_{\alpha}/2
  \end{pmatrix} \; ,
\end{equation}
where $\tilde{\varepsilon}_{\alpha} = \varepsilon_{\alpha} - \mu$ and
$\tilde{\omega}_{\vect{p}} = \omega_{\vect{p}} - \mu$ are respectively
the excitonic and photonic energies measured with respect to the
chemical potential. Within this path integral formulation, it is
possible to show~\cite{popov_fedotov} that the constraint on the site
occupation can be taken into account elegantly by shifting the
fermionic Matsubara frequencies according to
\begin{equation*}
  \epsilon_n = (2n + 1)\pi/\beta \mapsto \epsilon_n = (2n +
3/2)\pi/\beta \; ,
\end{equation*}
which we will assume from here on in.

\subsection{Mean-Field Phase Diagram}
\label{sec:phase}
We will see later in section~\ref{sec:photo} how the formation of
polaritons, the result of strong coupling between photons and
excitons, influences the optical response and in particular how the
optical response is affected when the system enters a condensed phase.
This section will discuss the mean-field equations, and resultant
phase diagram; however first we wish to stress an important point.  As
observed previously, the relevant polariton states form from the
mixing of a photon with a given momentum to many localised excitonic
states with different energies.  The strength of this mixing is
determined by the distribution of oscillator strengths.  Although the
quantum well excitons may be disorder-localised, for a weak disorder
potential, with $\sigma < \Omega_R$, and in the absence of strong
photon disorder (which in some cases can also be
relevant~\cite{langbein,gurioli}) the resultant polaritons are
delocalised. As such, polaritons will be described by a momentum
quantum number, and condense, as in the usual picture, in the lowest
momentum state.

The static and uniform minimum $\psi_{\vect{p}} (\tau) = \psi
\delta_{\vect{p} , 0}$ of the action~\eqref{eq:actio} can be found
from the equation of motion, or saddle-point equation, which has to be
solved together with the mean-field equation for the total number of
excitations~\eqref{eq:numbe}:
\begin{gather}
\label{eq:homo1}
  \tilde{\omega}_0 = \Frac{1}{N} \sum_{\alpha} |g_{\alpha,0}|^2
  \frac{\tanh \beta E_{\alpha}}{2 E_{\alpha}}\\ \rho = \frac{1}{N}
  \psi^2 + \frac{1}{N} \sum_{\alpha} \left[\frac{1}{2} -
  \frac{\tilde{\varepsilon}_{\alpha} \tanh \beta E_{\alpha}}{4
  E_{\alpha}}\right]\; .
\label{eq:homo2}
\end{gather}
Here, the spectrum of particle-hole excitations, i.e. the
quasi-particle spectrum discussed later, is controlled by the energy
of an exciton in presence of a coherent photon field $\psi$, and is
given by:
\begin{equation}
  E_{\alpha} = \sqrt{(\tilde{\varepsilon}_{\alpha}/2)^2 +
    |g_{\alpha,0}|^2 \psi^2/N} \; .
\label{eq:quasi}
\end{equation}

In the following, we will solve simultaneously these coupled
equations, both above ($\psi= 0$) and below ($\psi \ne 0$) the
critical temperature for condensation. It is interesting to compare
these with similar equations which have been already considered, e.g.,
in the BCS theory of superconductivity,~\cite{AGD} or in models
describing gases of fermionic atoms close to a Feshbach
resonance.\cite{FeshbachModels} Considering particularly the case of
Feshbach resonance, the density of particles is typically kept fixed,
while the interaction strength can be varied externally to span across
a BEC condensate of tightly bound fermionic molecules and a BCS state
of loosely bound fermionic pairs. In our case, two main important
differences arise. Firstly, density is the parameter which is changed.
Secondly, the coupling strength of excitons with cavity light depends
on the excitonic energy. There is a wide variety of behaviour that can
come from these coupled equations~\eqref{eq:homo1}
and~\eqref{eq:homo2}, controlled by the distribution of oscillator
strengths and by changing the density of states.

By solving the `gap equation' Eq.~\eqref{eq:homo1} in the limit $\psi
\to 0$, let us concentrate for the moment on the critical temperature
$T_c$ as a function of the chemical potential $\mu$, forgetting about
the corresponding excitation density. At the phase boundary, where
$\psi = 0$, in contrast to inside the condensed phase, the squared
oscillator strength appears only linearly in Eq.~\eqref{eq:homo1},
this expression therefore involving only the optical density
$D(\varepsilon)$. It is clear that the standard BCS result can be
thought of as a limiting case where the effective optical density is
much wider than the temperature.~\cite{peter_rev} In the opposite
limit, of a narrow delta-like optical density, one can solve the gap
equation analytically.~\cite{paul}

\begin{figure}
\begin{center}
\includegraphics[width=1\linewidth,angle=0]{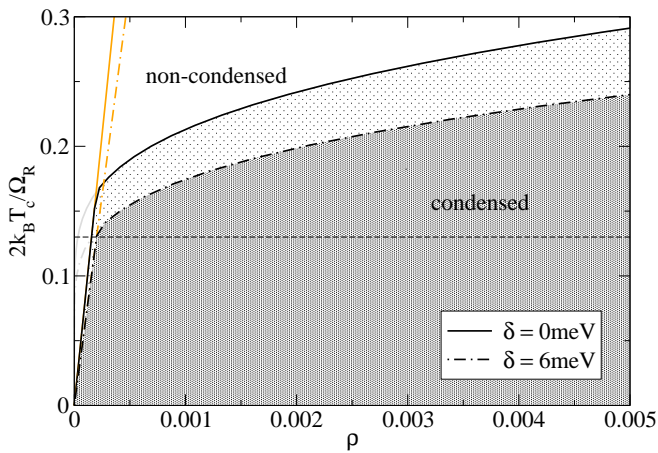}
\end{center}
\caption{\small (Color online) Phase diagram for the dimensionless
  critical temperature $2 k_B T_c /\Omega_{\mathrm{R}}$ versus the
  dimensionless density $\rho$ for effective zero detuning
  $\delta=\omega_0 - \varepsilon^*=0$meV ($\varepsilon^*$ is the
  energy at which the excitonic optical density has its maximum, and
  so as seen in Fig.~\ref{fig:optic}, $\delta=0$ implies $\omega_0 -
  E_{\ex} = -0.94$meV) (solid black) and for positive detuning
  $\delta=+6$meV ($\omega_0 - E_{\ex} = +5.06$meV) (dotted-dashed
  black) and $\Omega_{\mathrm{R}}=26$meV. The mean-field boundaries
  for the two different detunings are cut-off by the expected linear
  dependence of the critical temperature, as indicated at extremely
  low densities [orange (gray) solid and dotted-dashed lines].  The
  horizontal dashed line marks the temperature of $k_BT=20$K which
  will be used in later figures.}
\label{fig:criti}
\end{figure}

By taking into account the density equation~\eqref{eq:homo2}, the
critical temperature can be expressed as a function of the excitation
density rather than as a function of the chemical potential. In the
standard BCS theory there is no modulation of the interaction strength
and therefore no difference between optical density $D(\varepsilon)$
and density of states $\text{DoS} (\varepsilon)$. In our system
instead one has to introduce separately both quantities, and further a
realistic description of the DoS is required in order to measure the
density in physical units. In previous treatments,~\cite{paul} where
the DoS has been attributed an arbitrary Gaussian-like shape (and
moreover where the coupling strength was kept fixed), the density was
measured in arbitrary units. Note however that, the fact that the
density of states asymptotes to a constant at large densities, and the
presence of those exciton states that couple only weakly to light, is
responsible only for the asymptotic form of the phase boundary at
large densities (well beyond the range shown in Fig.~\ref{fig:criti}
and beyond the validity of our model). At smaller densities, such states have
a little effect on the phase boundary. In order to calculate the phase
boundary using our numerically calculated distributions, we will solve
Eq.~(\ref{eq:homo1}) and Eq.~(\ref{eq:homo2}) numerically.  How the
disorder averaged sums in these equations are to be evaluated has been
discussed Sec.~\ref{sec:numer}.

The critical temperature $T_c$ at which the system, for a given
density, goes from a non-condensed phase to a condensed phase, is
plotted in Fig.~\ref{fig:criti} for two different values of the
detuning $\delta = \omega_0 - \varepsilon^*$, where $\varepsilon^*$ is
the energy at which the excitonic optical density has its maximum (see
Fig.~\ref{fig:optic}). It is clear that increasing the detuning of the
cavity mode above the excitonic one decreases the overall photonic
fraction and therefore decreases the value of the critical temperature
at a given density of excitations. Conversely, if one fixes
temperature, the critical density for condensation is higher for
larger detuning, as Fig.~\ref{fig:criti} clearly shows. This
dependence is easy to understand in the extremely low density limit,
where the critical temperature varies linearly with density, and is
inversely proportional to the polariton mass: By increasing the
detuning, the polariton mass also increases.

Having discussed the thermodynamics via the mean-field equations, we
now turn in the rest of this paper to the analysis of the optical
responses and probes of microcavities, both below and above threshold
for condensation, with particular emphasis on the resonant Rayleigh
scattering emission. Such optical responses are described by
considering fluctuations about the mean-field theory outlined in this
section. The spectrum of these fluctuations, and the various optical
response functions that these fluctuations describe is the subject of
Sec.~\ref{sec:photo}.

As well as describing the optical response, fluctuations about the
mean field theory can also be important in describing, in the
extremely low density limit, corrections to the mean field
thermodynamics. Since the subject of this paper is optical probes, we
will not discuss the subject of fluctuation corrections to the
critical temperature here, as they have been discussed
elsewhere~\cite{jonathan}.

For low densities, a bosonic form can be recovered from our
Hamiltonian by using the Holstein-Primakoff transformation to express
the two-level systems in terms of bosonic operators, and so this low
density limit can of course be treated within our model.  A smooth
crossover between the low density (fluctuation dominated), and the
higher density (long-range interaction dominated) regimes can be found
by considering fluctuation corrections to the density, as was studied
in Ref~\onlinecite{jonathan}.

\section{Optical Probes}
\label{sec:photo}
The optical absorption and emission spectra of a microcavity can be
derived by first finding the Green's function describing photon
propagation.  The non-condensate part of these response functions
(i.e. away from zero momentum) may be found in practice by considering
the Green's function for fluctuations about the mean-field solution,
$\psi_{\vect{p},\omega_h} = \psi\, \delta_{\vect{p},0}
\delta_{\omega_h,0}+ \delta \psi_{\vect{p},\omega_h}$, where $\omega_h
= 2\pi h/\beta$ is a bosonic Matsubara frequencies, and expanding the
action~\eqref{eq:actio} up to quadratic terms:
\begin{equation*}
  \delta S \simeq \Frac{\beta}{2} \sum_{\omega_h , \vect{p}, \vect{q}}
  \begin{pmatrix}
    \delta \psi_{\omega_h , \vect{p}}^*\\ \delta \psi_{-\omega_h ,
    -\vect{p}}
  \end{pmatrix}^{T}
  \mathcal{G}^{-1}_{\vect{p}\vect{q}} (\omega_h)
  \begin{pmatrix}
    \delta \psi_{\omega_h , \vect{q}} \\ \delta \psi_{-\omega_h ,
    -\vect{q}}^*
  \end{pmatrix} \; .
\end{equation*}
This gives the Matsubara inverse photonic Green's function:
\begin{equation*}
  \mathcal{G}^{-1}_{\vect{p}\vect{q}} (\omega_h)= \begin{pmatrix}
  K^{(1)}_{\vect{p}\vect{q}} (\omega_h) & K^{(2)}_{\vect{p}\vect{q}}
  (\omega_h)\\
  K^{(2)}_{\vect{p}\vect{q}} (\omega_h) & {K^{(1)
  *}_{\vect{q}\vect{p}}} (\omega_h)
  \end{pmatrix}.
\end{equation*}
Physical response functions can be found by analytic continuation of
the imaginary time Matsubara Green's function to real
times.~\cite{AGD} The matrix elements of
$\mathcal{G}^{-1}_{\vect{p}\vect{q}} (\omega_h)$ can be expressed in
terms of the bare photon energy $\omega_{\vect{p}}$ and the excitonic
quasi-particle energy $E_{\alpha}$ as follows,
%
\begin{widetext}
\begin{align}
\label{eq:kern1}
  K^{(1)}_{\vect{p}\vect{q}} (\omega_h) &= \delta_{\vect{p},\vect{q}}
  \left(i \omega_h + \tilde{\omega}_{\vect{p}}\right) + \Frac{1}{N}
  \sum_{\alpha} g_{\alpha, \vect{p}}^* g_{\alpha, \vect{q}}\Frac{\tanh
  (\beta E_{\alpha})}{E_{\alpha}} \Frac{i \omega_h
  \tilde{\varepsilon}_{\alpha}/2 - E_{\alpha}^2
  -(\tilde{\varepsilon}_{\alpha}/2)^2}{\omega_h^2 + 4E_{\alpha}^2} -
  \delta_{\omega_h,0} \Frac{1}{N} \sum_{\alpha} \gamma_{\alpha ,
  \vect{p} , \vect{q}} \\
  K^{(2)}_{\vect{p}\vect{q}} (\omega_h) &= \Frac{1}{N} \sum_{\alpha}
  |g_{\alpha,0}|^2 g_{\alpha, \vect{p}}^* g_{\alpha, \vect{q}}
  \Frac{\psi^2}{N} \frac{\tanh (\beta E_{\alpha})}{E_{\alpha}}
  \frac{1}{\omega_h^2 + 4 E_{\alpha}^2} - \delta_{\omega_h,0}
  \Frac{1}{N} \sum_{\alpha} \gamma_{\alpha , \vect{p},\vect{q}}\; ,
\label{eq:kern2}
\end{align}
\end{widetext}
%
where
\begin{displaymath}
\gamma_{\alpha ,\vect{p}\vect{q}} = \beta |g_{\alpha,0}|^2
g_{\alpha, \vect{p}}^* g_{\alpha, \vect{q}} \frac{\psi^2}{N} 
\frac{\sech^2 (\beta E_{\alpha})}{4 E_{\alpha}^2}.
\end{displaymath}

It is useful to decompose the photonic Green's function into its
momentum diagonal and off-diagonal contributions:
\begin{equation}
  K_{\vect{p} \vect{q}}^{(1,2)} (\omega_h) =
  K_{\vect{p}\vect{p}}^{(1,2)} (\omega_h) \delta_{\vect{p} ,
  \vect{q}}+ K^{(1,2)o}_{\vect{p}\vect{q}} (\omega_h)\; .
\label{eq:offdi}
\end{equation}
By treating the off-diagonal terms perturbatively, the translational
invariance can be recovered and polariton eigenstates can be labelled
by momentum vectors. In the next section, we will focus on the
diagonal terms, which characterise both the spectral weight and the
photoluminescence emission. Section~\ref{sec:rrsca} is then dedicated
to RRS, for which we will see that the off-diagonal terms are
necessary in order to describe the scattering of an incident photon
(via the excitonic component of the microcavity polariton mode) into
directions other than its original direction.~\cite{whittakerRRS}

\subsection{Spectral Weight and Photoluminescence}
\label{sec:spect}
Secondary emission from a semiconductor microcavity after optical
excitation is the source of both incoherent PL and coherent RRS.  At
short times, this emission is dominated by RRS, the coherent
scattering from disorder, and so is at the energy of, and coherent
with, the incident radiation.  At longer times, phonon and
particle-particle scattering destroy coherence, and redistribute the
energy, leading to a quasi-equilibrium distribution of energies, and
thus the incoherent PL emission intensity,
\begin{equation}
  P(\omega , \vect{p}) = n_{B} (\omega) W(\omega,\vect{p})\; ,
\label{eq:phlum}
\end{equation}
is given by the Bose occupation factor $n_B (\omega)$ times the
spectral weight:
\begin{equation}
  W(\omega , \vect{p}) = 2 \left. \Im
  \mathcal{G}^{11}_{\vect{p}\vect{p}} (\omega_h) \right|_{i\omega_h =
  -\omega - i\eta} \; .
\end{equation}
The spectral weight can be interpreted as an absorption
coefficient~\cite{jonathan} (the probability to absorb a photon minus
the probability to emit a photon), where negative values of $W(\omega
, \vect{p})$ represent gain. In contrast, the PL $P(\omega ,
\vect{p})$ is always positive.

In calculating the PL, it is convenient to make an approximation by
neglecting multiple polariton scattering, while still including the
effects of exciton-disorder scattering to all orders. This is
discussed in Ref.~\onlinecite{whittaker}, where comparison between
this approximation and exact numerical calculations show this
approximation to be remarkably good. Physically, this is a good limit
to consider because the typical exciton-disorder scattering times are
very short compared to the inverse frequencies considered in PL. As PL
depends on the momentum diagonal part of the photon Green's function,
neglect of multiple scattering means in practice averaging over
disorder realisations at the level of the inverse photon Green's
function, $\mathcal{G}^{-1}_{\vect{p}\vect{q}} (\omega_h)$.  Since
off-diagonal terms in the inverse Green's function,
Eq.~\eqref{eq:offdi}, break translational invariance, they average to
zero, and so scattering between different photon momentum states can
thus be neglected. The off-diagonal terms neglected here will however
play a crucial role in the case of RRS response, as discussed below.

\begin{figure}
  \begin{center}
    \includegraphics[width=1\linewidth,angle=0]{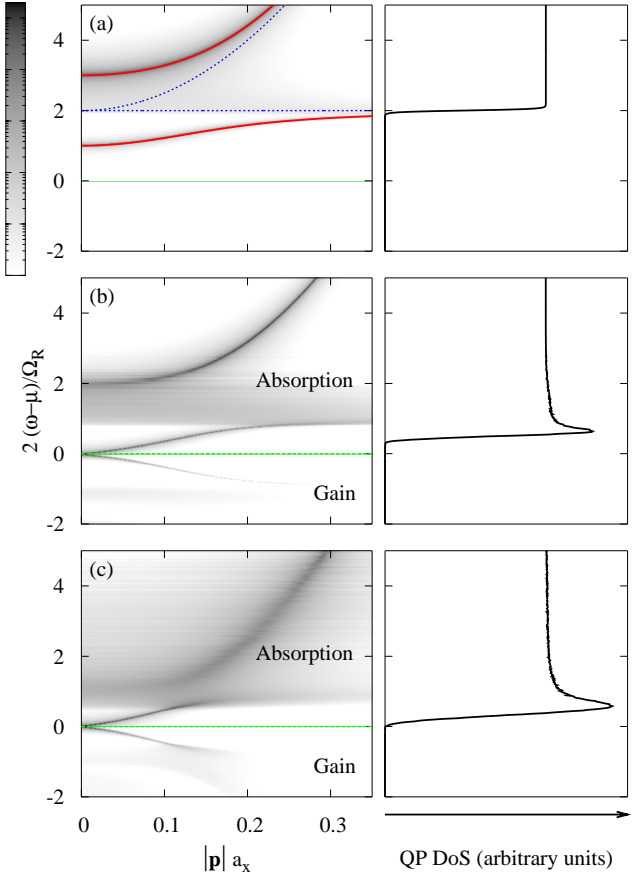}
  \end{center}
  \caption{\small (Color online) Left column: Contourplot of the
    spectral weight $W(\omega , \vect{p})$ as a function of the
    dimensionless momentum $|\vect{p}| a_{\ex}$ and rescaled energy
    $2(\omega - \mu)/\Omega_{\mathrm{R}}$, for zero detuning
    $\delta=\omega_0 - \varepsilon^*$ = 0 ($\omega_0 - E_{\ex} =
    -0.94$meV), $\Omega_{\mathrm{R}}=26$meV, $k_{B} T = 20$K and
    photon mass $m_{\text{ph}} \Omega_{\mathrm{R}} a_{\ex}^2/2 =
    0.01$.  Right column: plot of the quasi-particle DoS for the same
    choice of parameters as the left column, respectively: (a)
    non-condensed ($\rho\simeq 0$, $E_{\ex} - \mu \simeq 27$meV) (the
    bare exciton and photon dispersions [blue (dark gray) dotted line]
    and the upper and lower polariton dispersions [red (gray) solid
    line] obtained from the effective coupled oscillator model are
    shown for comparison); (b) condensed ($\rho\simeq 7.8\times
    10^{-3}$, $E_{\ex} - \mu \simeq 11$meV); (c) condensed
    ($\rho\simeq 6.7\times 10^{-2}$, $E_{\ex} - \mu \simeq 7$meV).}
  \label{fig:spwei}
\end{figure}

The spectral weight calculated from this formula, along with the
quasi-particle density of states is shown in Fig.~\ref{fig:spwei}.
We first discuss this in the non-condensed case, where $\psi=0$, and
so $K^{(2)}_{\vect{p}\vect{q}} (\omega_h) = 0$, and the photon's Green
function becomes diagonal in particle-hole space, and simplifies to:
\begin{align*}
  \mathcal{G}^{-1}_{\vect{p}\vect{p}} (\omega_h) &\simeq 
  i \omega_h + \tilde{\omega}_{\vect{p}} - 
  \int_{-\infty}^{\infty} d\varepsilon
  D(\varepsilon)
  \Frac{\tanh (\beta \tilde{\varepsilon}/2)}{i \omega_h
  + \tilde{\varepsilon}} \; .
\end{align*}
This expression describes the coupling of one harmonic oscillator (the
photon mode) to many harmonic oscillators (the exciton modes).  In the
limit of small density, the chemical potential is far below all
exciton modes, and so $\tanh(\beta\tilde{\varepsilon}/2) \simeq 1$.
In this limit, the result is identical to a bosonic description of
excitons sometimes also called the \emph{linear dispersion model}.
The underlying fermionic structure appears as a reduction of the
effective exciton-photon coupling, due to saturation effects described
by the $\tanh(\beta\tilde{\varepsilon}/2)$ term, and discussed in more
detail below.

From this \emph{linear dispersion model}, describing one oscillator
coupled to many, there are in general two broadened modes at high and
low energies -- here these are the lower polariton (LP) and upper
polariton (UP) -- and a continuum of modes associated with the exciton
optical density.  However, when considering the corresponding spectral
weight, modes are weighted by their photonic component, so one
primarily sees the LP and UP modes, and only weak emission near the
bare excitonic states -- the excitonic `dark' states.  These three
features are clearly visible in Fig.~\ref{fig:spwei}, and have been
previously predicted by Houdr\'e \emph{et al.}~\cite{houdre} making
use of a simplified and exactly solvable model. Note that in the limit
of an infinitely narrow spectral width, these dark exciton states
become entirely dark, having vanishing photon component. The presence
of `dark' exciton states, coexisting with strong coupling polaritons
can explain the simultaneous observation of large Rabi splittings and
long decay times seen in some experiments.~\cite{mintsev}

When the Rabi splitting is substantially larger than the exciton
inhomogeneous broadening, (in our case we have $\Omega_R=26$meV and
the FWHM of the optical density $\sigma^*=0.94$meV), there is a
substantial difference between broadening of lower and upper
polaritons.  This is because the high energy tail of the optical
density decays as a power law, while the low energy Lifshitz tail
decays exponentially. Thus, the optical density has a larger value at
the UP mode than at the LP mode, giving a larger broadening compared
to the almost vanishing width of the LP. This description of the
polariton linewidth due to the excitonic inhomogeneous broadening
coincides with that of Whittaker~\cite{whittaker}, and it has been
well tested experimentally.~\cite{borri}

In addition, the location of the LP and UP can be found by making use
of an effective two-oscillator model, i.e. assuming a narrow
delta-like optical density,
\begin{equation*}
  D(\varepsilon) \mapsto \left(\Frac{\Omega_R}{2}\right)^2
  \delta(\varepsilon - \varepsilon^*) \; ,
\end{equation*}
where $\varepsilon^*$ is the location of the maximum optical density
and effectively, the exciton energy. In this case the system reduces
to two coupled oscillators, giving unbroadened LP and UP poles at
\begin{equation}
  E_{\text{LP,UP}} = \Frac{\tilde{\omega}_{\vect{p}} +
  \tilde{\varepsilon}^*}{2} \pm \frac{1}{2}
  \sqrt{\left(\tilde{\omega}_{\vect{p}} -
  \tilde{\varepsilon}^*\right)^2 + \overline{\Omega}_R^2}\; ,
\label{eq:lpeup}
\end{equation}
where , $\overline{\Omega}_R^2 \equiv \Omega_R^2 \tanh(\beta
\tilde{\varepsilon}^*/2)$ is the reduced Rabi splitting, due to
saturation effects at higher densities. The reduction of
$\overline{\Omega}_R$ splitting thus translates directly into a
reduction of the LP -- UP splitting, and thus a blue-shift of the LP,
which at small densities can be shown to be linear. For comparison,
the results of this formula are shown by the solid (red) lines in
panel (a) of Fig.~\ref{fig:spwei}. In Sec.~\ref{sec:blues} we will
discuss the calculation of this linear blue-shift in the low density
regime, by evaluating $\overline{\Omega}_R$ as a function of density,
making use of the full density of states.

Let us now turn to the signatures of condensation as seen in the
spectral weight, and thus in the PL emission (but which are most
probably masked in the PL emission by strong emission from the
condensate mode). When condensed, the polariton modes are replaced by
new collective modes:~\cite{jonathan,RRS_letter} The lower polariton
becomes the linear Goldstone mode, and two branches appear below the
chemical potential.
The appearance of new excitation branches below the chemical potential
is generic to condensation,~\cite{StringariBook} however the
experiments required to probe these modes are not easy in other Bose
condensed systems, such as atomic gases.
For this reason, let us briefly discuss the physical origin of these
new branches, and the reason they may be observed in optical response
of polariton systems.
The Bogoliubov spectrum arises because of the possibility of processes
that spontaneously either create or destroy two non-condensed
particles.  (Such processes arise due to scattering from or to the
condensate.)  As a result of these processes, there is mixing between
the propagation of an extra particle, or propagation of a missing
particle (i.e. a hole).  (Such a language of particle and hole refers
to the normal state quasi-particles, in the current case the normal
state polaritons.)  In the normal state, one can separately calculate
the spectral weight of particle excitations, which have weight only
above the chemical potential, and hole excitations which have weight
only below the chemical potential. When condensed, as the ``particle''
and ``hole'' spectral weights become mixed, this mixing can lead to
spectral weight below the chemical potential that is associated with
particle propagation. To observe this weight it is however necessary
to have the ability to inject a particle which is not a quasi-particle
of the condensed system, i.e. not a Bogoliubov quasiparticle.  In
atomic experiments this is hard to achieve, but for polaritons can be
naturally achieved by injecting a photon.
These new branches below the chemical potential are seen as optical
gain in the spectral weight (see Fig.~\ref{fig:spwei}).
Note that the presence of pumping and decay will modify the linear
dispersion of the Goldstone mode at low momentum, making it
diffusive,~\cite{marz_keldysh} and that quantisation by disorder may
also have some effect.~\cite{alexy_alexy}

\begin{figure}
\begin{center}
\includegraphics[width=1\linewidth,angle=0]{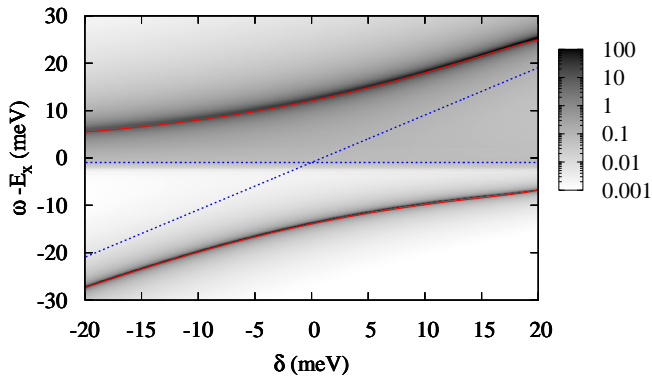}
\end{center}
\caption{\small (Color online) Contourplot of the spectral weight
  $W(\omega,0)$ for $\vect{p}=0$ as a function of the detuning
  $\delta=\omega_0 - \varepsilon^*$ and energy $\omega-E_{\ex}$, for a
  fixed and very low value of density ($\rho \simeq 0$, $E_{\ex} - \mu
  \simeq 27$meV) and $k_BT = 20$K ($2k_BT/\Omega_R = 0.13$). The bare
  exciton and photon dispersions [blue (dark gray) dotted line] and
  the upper and lower polariton dispersions [red (gray) solid line]
  obtained from the effective coupled oscillator model are explicitly
  shown.}
\label{fig:vsdet}
\end{figure}

The spectral weight also contains information about the excitonic
quasi-particle DoS of the system.  As discussed above for the
non-condensed case, this is visible via the appearance of `dark'
exciton states [see Figs.~\ref{fig:vsdet} and~\ref{fig:spwei}(a)].
When condensed, there is a coherent field that modifies the energies
of these excitonic quasi-particle states, as given in
Eq.~(\ref{eq:quasi}), and thus modifies their density of states.  The
density of states given by taking the energy of such modes is shown by
the right column of Fig.~\ref{fig:spwei}, and can be compared to the
corresponding faint features seen in the photon spectral weight in the
left column.

The change to the quasi-particle density of states that occurs on
condensation requires some explanation.  As mentioned earlier, the
change of this spectrum is similar to that seen in the BCS theory of
superconductivity, as such it is surprising that there is no gap in
the density of states of Fig.~\ref{fig:spwei}.  In fact, there would
be a gap if the distribution of oscillator strengths were replaced
with the mean square oscillator strength.  In that case, there would
be a branch cut, and a $(E-g\psi/\sqrt{N})^{-1/2}$ singularity in the
density of states.  However, with the full distribution of oscillator
strengths, one finds that there is always a non-vanishing probability
of arbitrarily small oscillator strengths.  (Note that if the
oscillator strength were zero, the corresponding state would not
contribute to any photon response, but for an arbitrarily small
coupling, it has some contribution.)  Thus, the contribution of these
weakly coupled exciton states in effect smoothes out the gap.  Thus,
this system can be vaguely described as `gapless fermion
condensation', analogous to `gapless superconductivity', but through a
mechanism very different to the standard Abrikosov-Gor'kov mechanism
considered in superconductors.~\cite{AGD,marzena}

As one moves away from the chemical potential, the mixing of
`particle' and `hole' modes described by the Bogoliubov spectrum
reduces.  As a result, far above the chemical potential, the modes are
much as the uncondensed case, and the far below the chemical
potential, the new modes disappear.  To explain this quantitatively,
it is clearer to discuss the case of weakly interacting Bose
gas.~\cite{StringariBook} Writing the Bogoliubov mode energy as
$\xi_{\vect{p}} = \sqrt{\epsilon_{\vect{p}} (\epsilon_{\vect{p}} +
2\mu)}$, where $\mu = g\psi^2$ is the mean-field value, the spectral
weight is given by:
\begin{equation*}
  W(\omega , \vect{p}) = \Frac{\epsilon_{\vect{p}} + \mu +
  \xi_{\vect{p}}}{2 \xi_{\vect{p}}} \delta (\omega - \xi_{\vect{p}}) -
  \Frac{\epsilon_{\vect{p}} + \mu - \xi_{\vect{p}}}{2 \xi_{\vect{p}}}
  \delta (\omega + \xi_{\vect{p}}) \; ,
\end{equation*}
It is thus clear that at large momenta, where $\xi_{\vect{p}} \simeq
\epsilon_{\vect{p}} + \mu$, the coefficient of the first delta
function (modes above the chemical potential) will go to one, and the
the coefficient of the second (modes below) will be suppressed to zero
(roughly quadratically in energy).

\begin{figure}
\begin{center}
\includegraphics[width=1\linewidth,angle=0]{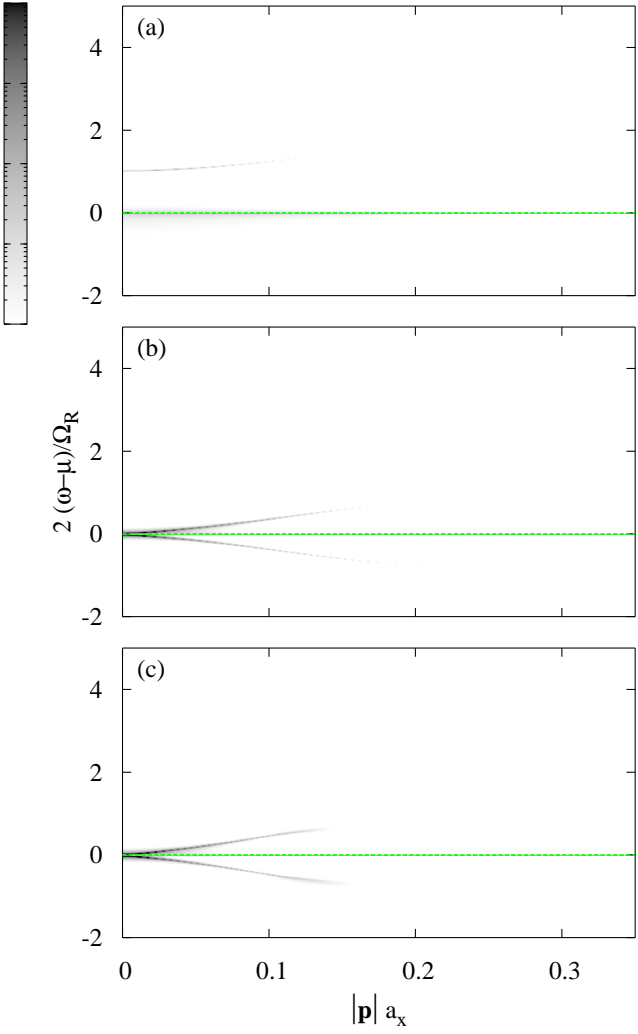}
\end{center}
\caption{\small (Color online) Contourplot of the incoherent PL $P
  (\omega , \vect{p})$ as a function of the dimensionless momentum
  $|\vect{p}| a_{\ex}$ and rescaled energy $2(\omega -
  \mu)/\Omega_{\mathrm{R}}$, for detuning $\omega_0 - E_{\ex} =
  -0.94$meV, $\Omega_{\mathrm{R}}=26$meV, $k_{B} T = 20$K and
  $m_{\text{ph}} \Omega_{\mathrm{R}} a_{\ex}^2/2 = 0.01$: (a)
  non-condensed ($\rho\simeq 0$, $E_{\ex} - \mu \simeq 27$meV); (b)
  condensed ($\rho\simeq 7.8\times 10^{-3}$, $E_{\ex} - \mu \simeq
  11$meV); (c) condensed ($\rho\simeq 6.7\times 10^{-2}$, $E_{\ex} -
  \mu \simeq 7$meV).}
\label{fig:plumi}
\end{figure}

In contrast to the power law suppression of the spectral weight of
modes far below the chemical potential, the PL signal from modes far
above zero is suppressed exponentially by the thermal occupation of
these modes [see Eq.~\eqref{eq:phlum}], while below the chemical
potential, there is no such decay. However, this discussion neglects
emission from the condensate mode, which should included at zero
momentum, as defined in Eq.~\eqref{eq:phlum}. To see a noticeable
change of the normal modes requires a relatively large condensate
density. Due to instrumental broadening, the presence of emission from
this large condensate density might obscure emission from the normal
modes, so they may only be weakly visible, as shown in
Fig.~\ref{fig:plumi}. As discussed below, RRS may provide a means of
escape from this problem.
%

\subsection{Lower Polariton Blue-Shift}
\label{sec:blues}
Before discussing the RRS response, let us briefly discuss a point
mentioned above -- the calculation of the LP blue-shift as a function
of density, making use of the full density of states, and our model of
saturation effects~\eqref{eq:hamil}. This can be observed in PL
experiments by the change of blue-shift as a function of intensity of
the non-resonant pumping.

In the non condensed regime, the blue-shift of the LP in our model is
a consequence of the saturation of the (disordered) energy levels.
Following the narrow band-width limit described in
Eq.~(\ref{eq:lpeup}), this saturation blue-shift can be found if one
has an expression for the chemical potential at a given temperature
and density.  In the low density limit, the expression for density
$\rho$ as a function of $\mu$ [Eq.~(\ref{eq:homo2}) with $\psi=0$] can
be inverted in terms of elementary functions.  As illustrated by
Fig.~\ref{fig:occup}, in the low density limit, the chemical potential
is far below the band edge. Thus, there are two significant
contributions to density: one from low energies, in the tail of the
DoS, but at large occupation, and one from high energies, in the tail
of the occupation, but at large DoS, thus:
\begin{equation}
  \label{eq:density-assympt}
  \rho \simeq \int_{-\infty}^{\mu} \Frac{d\varepsilon}{\Ryx} e^{-
  2|\varepsilon|/W_{\rho}} + \int_{0}^{\infty}
  \Frac{d\varepsilon}{\Ryx} e^{-(\varepsilon - \mu)/k_B T}\; .
\end{equation}
Here $W_{\rho}/2 = 0.32$meV is the energy which characterises the
exponential decay of DoS in the tail of the Lifshitz states -- this
coefficient is extracted by an exponential fit to our numerical DoS.
At a temperature of $k_BT = 20K = 1.72$meV, the dominant contribution
to the density is given by the second integral and the chemical
potential increases logarithmically with the density:
\begin{equation}
  \rho \simeq \Frac{k_BT}{\Ryx} e^{\mu/k_B T}\; .
\label{eq:lowde}
\end{equation}
From this expression, and from the coupled oscillator expressions of
the LP Eq.~\eqref{eq:lpeup}, we can explicitly derive the reduced Rabi
splitting due to saturation effects,
\begin{equation*}
  \overline{\Omega}_R \simeq \Omega_R \left(1 - 2e^{-\varepsilon^*/k_B
  T} \Frac{\Ryx}{k_B T} \rho\right)^{1/2}
\end{equation*}
and the LP blue-shift, which is thus linear in this low-density
regime:
\begin{multline}
  \delta E_{\text{LP}} \equiv E_{\text{LP}} (\rho) - E_{\text{LP}} (0)
  \sim \Omega_R n \Frac{1}{m_{\ex} k_B T} \; ,
\label{eq:appbl}
\end{multline}
where we have reintroduced here the density $n$ per unit volume.
Remaining at small densities, but now considering the low temperature
limit, $W_{\rho}/2 > k_B T$, then the dominant term in
Eq.~(\ref{eq:density-assympt}) is the first, and so
Eq.~(\ref{eq:appbl}) should be modified by replacing the thermal length
$(m_{\ex} k_B T)^{-1}$, by a characteristic disorder length $(m_{\ex}
W_{\rho})^{-1}$.

It is instructive to compare this result with that for a clean system;
in this case blue-shift of the LP has been attributed either due to
Coulomb interaction~\cite{rochat} or to saturation
effects~\cite{ciuti} (where the expression given here is valid only in
the dilute limit):
\begin{align*}
  \delta E^0_{\text{Coul}} &\sim \Ryx n a_{\ex}^2\\
  \delta E^0_{\text{sat}} &\sim \Omega_R n a_{\ex}^2\; .
\end{align*}
Because the excitonic Rydberg $\Ryx$ can be of the same order of
magnitude as the Rabi splitting $\Omega_R$, in a clean system the two
shifts can be expected to be of the same order of
magnitude. Considering just saturation effects, the difference between
clean and dirty systems is that, at low temperature, in a clean system
blue-shift depends on exciton number per square Bohr radius, while in
a disordered system the relevant length is that characteristic of the
disorder potential, which is in general larger than the Bohr radius.
Finally, we wish to observe that there is a distinction between
Coulomb and saturation effects: Coulomb interactions result in a
blue-shift of both the LP and UP, while saturation leads to a
blue-shift of the LP and a red shift of the UP, i.e. a collapse of the
Rabi splitting.  Thus, these effects can be experimentally
distinguished, and their relative magnitudes determined.

\begin{figure}
\begin{center}
\includegraphics[width=1\linewidth,angle=0]{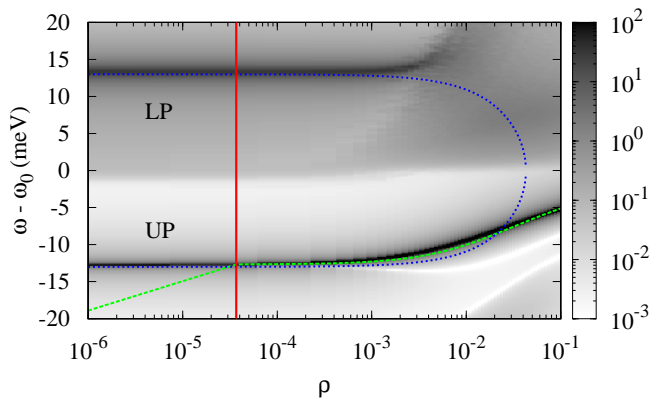}
\end{center}
\caption{\small (Color online) Contourplot of the spectral weight
  $W(\omega,0)$ for $\vect{p}=0$ as a function of the dimensionless
  density $\rho$ and the rescaled energy $\omega - \omega_0$, for zero
  detuning ($\omega_0 - E_{\ex} = -0.94$meV) and $k_BT = 20$K
  ($2k_BT/\Omega_R = 0.13$). The rescaled chemical potential
  $(\mu-\omega_0)$ [green (light gray) dashed], the non-condensed
  lower and upper polariton modes from the two coupled oscillator
  model [blue (dark gray) dotted], and the critical density for
  condensation $\rho_c \simeq 3.7\times 10^{-5}$ [vertical solid red
  (dark gray) line] are shown for comparison.}
\label{fig:vsden}
\end{figure}

In the condensed regime, the equivalent of the LP is the Goldstone
linear mode, which by definition starts at the chemical potential.
Thus, the observed blue-shift is a direct observation of chemical
potential vs.\ density. Figure~\ref{fig:vsden} shows the variation of
the spectral weight at zero momentum as a function of density, from
which the energy of the zero momentum LP and UP modes can be
extracted.  Also shown, for comparison, is the chemical potential vs.\
density (dashed green line).  The locking of the LP mode to the
chemical potential beyond the critical density (red solid line) is
clearly visible, and the behaviour of the LP mode vs density allows
the extraction of density from an experimental measurement of LP
blue-shift.
%

\subsection{Resonant Rayleigh Scattering}
\label{sec:rrsca}
As described in Refs.~\onlinecite{whittakerRRS,shchegrov}, the RRS
intensity describes the probability to inject a photon into the cavity
with momentum $\vect{p}$, and detect a coherent photon at $\vect{q}$
with $\vect{p}\ne\vect{q}$.  The intensity of such a signal may be
written as:
\begin{equation}
  I_{\vect{p}\vect{q}} (\omega) = \left. |\mathcal{G}^{11}_{\vect{p}
  \vect{q}}|^2 \right|_{i\omega_h = -\omega - i\eta} \; .
\label{eq:rrsin}
\end{equation}
Since this involves the off-diagonal momentum space components of the
Green's function, it is necessary to include the off-diagonal momentum
components of the inverse Green's function Eq.~\eqref{eq:offdi}.
However, as discussed in Sec.~\ref{sec:photo}, we will consider only a
single polariton scattering and again neglect multiple polariton
scattering (but include all orders of exciton-disorder scattering).
Such an approximation is reasonable for the same reasons discussed at
the start of Sec.~\ref{sec:photo}.

In the following, we propose using the RRS signal to probe the
excitation spectrum in the presence of a polariton condensate. It is
therefore necessary to be able to separate the RRS signal from the
strong photoluminescence that would arise from the equilibrium state
with large polariton density. Further, since a condensed polariton
system may have a strong nonlinear response to an applied probe
(i.e. stimulated scattering if the probe significantly affects the
population of polariton modes), it is necessary to use a weak RRS
probe. Fortunately, the coherent nature of RRS allows exactly this:
one can detect a weak RRS signal by phase sensitive measurement, as is
discussed in detail in Appendix~\ref{sec:phase-sens-detect}. Hence,
the limit on intensity of the probe is provided by the sensitivity of
the CCD camera, and not by the incoherent photoluminescence
background. This therefore allows a probe sufficiently weak that only
the linear RRS response is seen, and nonlinear effects can be avoided.

\begin{figure}
\begin{center}
\includegraphics[width=1\linewidth,angle=0]{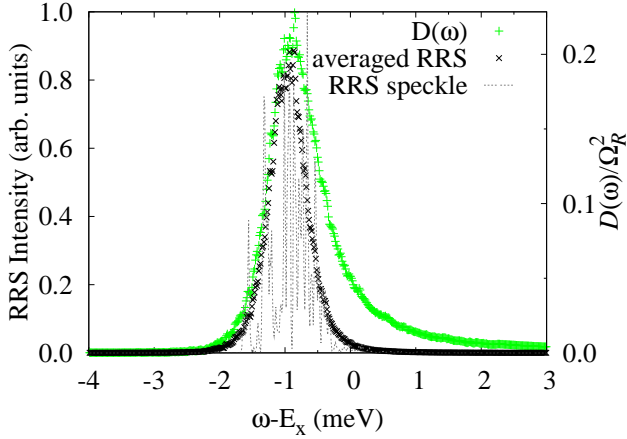}
\end{center}
\caption{\small (Color online) Excitonic speckle pattern
  $|K^{(1o)}_{\vect{p},\vect{q}} (\omega_h)|^2|_{i\omega_h = -\omega -
    i\eta}$ for a single disorder realization (gray dotted line) and
  its disorder average (black cross symbols) versus energy for
  $|\vect{p}| = |\vect{q}| = 36^{\circ}$ and a $90^{\circ}$ azimuthal
  angle, $\vect{p} = (0,6.3\times 10^4) \text{cm}^{-1}$ and $\vect{q}
  = (6.3\times 10^4,0) \text{cm}^{-1}$. For comparison, we also plot
  the excitonic optical density [green (light gray) plus symbols].}
\label{fig:excsp}
\end{figure}

When non-condensed, the single polariton scattering expansion of the
photon Green's function in the off-diagonal terms gives
\begin{equation}
  I_{\vect{p}\vect{q}} (\omega_h) \simeq
   \Frac{1}{|K^{(1)}_{\vect{p}\vect{p}} (\omega_h)|^2}
   |K^{(1)o}_{\vect{p}\vect{q}} (\omega_h)|^2
   \Frac{1}{|K^{(1)}_{\vect{q}\vect{q}} (\omega_h)|^2} \; .
\label{eq:expan}
\end{equation}
The factors $|K_{\vect{q}\vect{q}}^{(1)}(\omega_h)|^{-2}$ appearing
here can be interpreted as a filter, allowing a response only when the
outgoing (or equivalently incoming for $\vect{q} \to \vect{p}$)
momentum has an energy close to the polariton mode at the given
energy.  This means that $|\vect{p}| \simeq |\vect{q}|$ and so is
responsible for the ring-shaped RRS signal observed in
experiments.~\cite{freixanet,langbein} In contrast, the term
$|K^{(1o)}_{\vect{p},\vect{q}} (\omega_h)|^2$ describes scattering
between momentum states via polariton-exciton-polariton scattering.
This term has a large variation from disorder realization to disorder
realization, and is the reason for the speckle seen in RRS
experiments.  This speckle, and the disorder averaged RRS intensity
are shown for comparison in Fig.~\ref{fig:excsp}.  As the precise
speckle pattern depends on the precise disorder realization, the most
we can reasonably do is to describe the statistical properties of this
speckle, thus the disorder averaged RRS signal shown in
Fig.~\ref{fig:rrsca} would in experiment represent the envelope of the
RRS speckle pattern.

When condensed, the expression for the RRS intensity becomes more
complicated, but can still be written in the form of a filter and
scattering part:
\begin{equation}
  I_{\vect{p} \vect{q}} (\omega) = \simeq F_{\vect{p}} S_{\vect{p}
  \vect{q}} F_{\vect{q}} \; ,
\label{eq:filsc}
\end{equation}
where the filter and scattering terms are respectively:
\begin{equation}
\begin{split}
  F_{\vect{p}} &= \big| |K^{(1)}_{\vect{p} \vect{p}}|^2 -
  (K^{(2)}_{\vect{p} \vect{p}})^2 \big|^{-2},\\
  S_{\vect{p} \vect{q}} &= \bigg| K^{(1)o}_{\vect{p} \vect{q}} K^{(1)
  *}_{\vect{p} \vect{p}} K^{(1) *}_{\vect{q} \vect{q}} + K^{(1)o
  *}_{\vect{q} \vect{p}} K^{(2)}_{\vect{p} \vect{p}} K^{(2)}_{\vect{q}
  \vect{q}} \\
  &- K^{(2)o}_{\vect{p} \vect{q}} \left(K^{(2)}_{\vect{p} \vect{p}}
  K^{(1) *}_{\vect{q} \vect{q}} + K^{(1) *}_{\vect{p} \vect{p}}
  K^{(2)}_{\vect{q} \vect{q}}\right) \bigg|^2 \; .
\end{split}
\end{equation}
As in the non-condensed case, the filter function $F_{\vect{p}}$
restricts the allowed incoming and outgoing momenta to those for which
the normal modes have the injected energy.  However, as discussed
earlier, in the presence of a condensate, the normal modes supported
are no longer the LP and UP, but instead the new Bogoliubov-like
quasi-modes.

The decay of spectral weight associated with modes far below the
chemical potential, as discussed in Sec.~\ref{sec:spect} still applies
(this can clearly be seen in Fig.~\ref{fig:rrsca},
Fig.~\ref{fig:spwei} and Fig.~\ref{fig:plumi}).  However, since the
energy of the RRS signal is controlled by the energy of the incident
photon, it is possible to study modes far above the chemical
potential, which would have negligible equilibrium occupation, and
thus negligible weight in the PL signal. 
%
%
Both for this reason, and because of the ability to discriminate
between the coherent RRS signal and inchoherent photoluminescence, RRS
provides a very powerful tool to study the interesting properties of
condensed polariton systems.

\section{Conclusions}
\label{sec:concl}
In this paper we have analysed and compared different optical
responses of a condensate of microcavity polaritons. One of the
conclusions of this work is that, compared to photoluminescence
studies, the energy and momentum resolved resonant Rayleigh scattering
spectra provides many advantages in probing a condensed phase and in
studying the associated coherent excitation spectrum. In particular,
resonant Rayleigh scattering, by collecting the signal at a different
angle from that at which the system is probed, allows one to look
directly at characteristic features of the condensed phase excitation
spectrum. Such features, like the Goldstone modes both above and below
the chemical potential, are weak in photoluminescence emission and are
likely to be masked by the strong emission from the condensate in the
lowest momentum state. In contrast, Rayleigh scattering, being
coherent with the probe, can be distinguished from condensate
emission.

One can ask why the accurate description of excitonic disorder
considered in this paper is necessary, when the microcavity photon
introduces a long length scale which averages over disorder.  One
might also compare this system with other excitonic systems, such as
double quantum wells,~\cite{butov_review} where high density of
dipole-dipole interacting excitons quickly screens the
disorder.~\cite{alex_qw,zimmermann_qw} The answer to this question is
that the large ratio of exciton to photon mass means firstly that
exciton density required for condensation in microcavities is far less
than in double quantum wells; and secondly that those exciton states
involved in forming the thermally occupied polariton modes are
strongly localised exciton states (i.e. influenced by disorder beyond
Born approximation). One of the consequences of an accurate treatment
of disorder in microcavities is that the blue-shift of the lower
polaritons due to saturation effects depends on a length scale
characteristic of disorder and temperature, which can be much larger
than the Bohr radius, this length playing an equivalent role in the
clean system. This effect can be important in determining the
polariton densities from the measured blue-shift in current
experiments.

An important consequence of disorder is that the exciton-light
coupling strength is characterised by a full distribution. In the
non-condensed phase, the optical density alone determines the
photoluminescence response, while for resonant Rayleigh scattering the
averaged fourth power of the oscillator strength is required. However,
in the condensed phase one has to consider the entire distribution for
each exciton energy. For energies close to and above the band edge,
there is always a non-vanishing probability of an arbitrarily small
oscillator strength. For this reason, in contrast to examples like BCS
superconductivity, one can show that the quasi-particle spectrum does
not have a hard gap.

In conclusion, we have considered how an accurate treatment of
disorder on the single-particle excitonic level, when elevated to the
many body problem of interacting microcavity polaritons, leads to a
variety of interesting features in various optical responses and
probes of the condensed phase.

\begin{acknowledgments}
  We are grateful to Roland Zimmermann, Wolfgang Langbein and Ben
  Simons for suggestions and useful discussions. F.M.M. and M.H.S.
  would like to acknowledge financial support from EPSRC. J.K. would
  like to acknowledge financial support from the Lindemann Trust. This
  work is supported by the EU Network ``Photon mediated phenomena in
  semiconductor nanostructures'' HPRN-CT-2002-00298.
\end{acknowledgments}

\appendix
\section{Phase sensitive detection of resonant Rayleigh signal}
\label{sec:phase-sens-detect}
In this appendix we show how the coherent nature of the resonant
Rayleigh signal allows a weak coherent probe to be detected in the
presence of strong incoherent photoluminescence.
The Rayleigh scattering probe is a perturbation, $\hat{H} = \hat{H}_0
+ \hat{V}$, which in the rotating wave approximation may be written
as:
\begin{equation}
  \hat{V}=A_0 \left( \hat{\psi}^{\dagger}_{\vect{p}} e^{i\Omega t} +
    \hat{\psi}^{}_{\vect{p}} e^{-i\Omega t} \right)\; .
\label{eq:2}
\end{equation}
This describes a probe at wavevector $\vect{p}$, frequency $\Omega$,
of strength $A_0$.
The RRS signal is the coherent scattering of this probe to other
wavevectors.
To isolate the part of the emission that is coherent with the probe,
one may use a homodyne measurement, interfering the emission with part
of the probe signal.
This corresponds to measuring the spectrally resolved emission
intensity :
\begin{equation}
  P(\omega, \vect{q}) = \int dt e^{i\omega t} P(t,\vect{q}) \; ,
\label{eq:1a}
\end{equation}
where
\begin{multline*}
  P(t,\vect{q}) = \sum_n e^{\beta(F-E_n)} \\ 
  \times \langle n | \left[\hat{\psi}^{\dagger}_\vect{q}(t) +
  A_1e^{-i(\Omega t + \phi)}\right] \left[\hat{\psi}^{}_\vect{q}(0) +
  A_1e^{i\phi} \right] |n\rangle \; ,
\end{multline*}
and where $A_1$ is the strength of the homodyne mixing, and $\phi$ is
a phase delay introduced between the probe and the homodyne signal.
The states $|n \rangle$ are the eigenstates of the system without the
probe, given by $\hat{H}_0 | n \rangle = E_n | n \rangle$; $F$ is the
free energy, for normalization.

The principle of phase sensitive detection is to vary the phase delay,
and to extract the part of $P(\omega,\vect{q})$ which depends on this
phase delay.
Since the background PL does not depend on the phase delay, this
allows one to separate a small but phase dependent signal from a
strong but phase independent background.

The part of the signal that depends on phase is:
\begin{multline}
  P_{\phi}(t,\vect{q}) = A_1 \sum_n e^{\beta(F-E_n)} \\
  \times \left[ \langle n | \hat{\psi}^{\dagger}_\vect{q}(t) | n
  \rangle e^{i\phi} + \langle n | \hat{\psi}^{}_\vect{q}(0) | n
  \rangle e^{-i(\phi+\Omega t)} \right] \; .
\label{eq:4}
\end{multline}
Standard first order time-dependent perturbation theory, using the
perturbation in Eq.~\eqref{eq:2} yields:
\begin{multline}
  \langle n | \hat{\psi}^{\dagger}_\vect{q}(t) | n \rangle = A_0
  \sum_m \left[ \frac{ \langle n | \hat{\psi}^{\dagger}_\vect{q} | m
  \rangle \langle m | \hat{\psi}^{}_\vect{p} | n \rangle }{E_n - E_m -
  \Omega - i \eta} \right.  \\
  - \left.  \frac{ \langle n | \hat{\psi}^{}_\vect{p} | m \rangle
  \langle m | \hat{\psi}^{\dagger}_\vect{q} | n \rangle }{E_m - E_n -
  \Omega - i \eta} \right]e^{-i \Omega t} + [\ldots] e^{i\Omega t} \;
  .
\label{eq:5}
\end{multline}
The term written $[\ldots]$ is similar to the first term, but with
$\Omega \mapsto - \Omega$ and $\psi^{}_{\vect{p}} \mapsto
\psi^{\dagger}_{\vect{p}}$.
Inserting this signal into Eq.~\eqref{eq:4} and then into
Eq.~(\ref{eq:1a}), it can be seen that such a term gives a signal at
frequency $\omega=-\Omega$, and so can be clearly separated and
ignored.
Inserting the first term into Eq.~\eqref{eq:4}, one can write:
\begin{multline}
  \sum_n e^{\beta(F-E_n)} \langle n |
  \hat{\psi}^{\dagger}_\vect{q}(t)| n \rangle = \sum_{n,m}
  e^{\beta(F-E_n)} \\
  \times \left[1-e^{\beta(E_n - E_m)}\right] \frac{ \langle n |
  \hat{\psi}^{\dagger}_\vect{q} | m \rangle \langle m |
  \hat{\psi}^{}_\vect{p} | n \rangle }{E_n - E_m - \Omega - i \eta} \;
  ,
\label{eq:6} 
\end{multline}
which is the definition of the retarded Green's function,~\cite{AGD}
$G^{11}_{R,\vect{p}\vect{q}}(\Omega)=
\left.\mathcal{G}^{11}_{\vect{p}\vect{q}} \right|_{i\omega_h=-\Omega -
i \eta}$.
By repeating the same analysis for the second term in
Eq.~(\ref{eq:4}), one can then write:
\begin{equation}
  P_{\phi}(t,\vect{q}) = 2 A_1 A_0 \left|
  G^{11}_{R,\vect{p}\vect{q}}(\Omega) \right| \cos(\phi + \phi_0)
  e^{-i\Omega t} \; ,
\label{eq:7}
\end{equation}
where $\phi_0$ is the phase of the retarded Green's function.
Hence, the phase-dependent part of the luminescence is given by the
off-diagonal in momentum space part of the retarded Green's function,
i.e. the RRS signal.

%

\end{document}